\newcommand\apjcls{1}
\newcommand\aastexcls{2}
\newcommand\othercls{3}
\newcommand\papercls{\aastexcls}

\documentclass[tighten, times, twocolumn, twocolappendix]{aastex63}  

\if\papercls \apjcls
\usepackage{apjfonts}
\else\if\papercls \othercls
\usepackage{epsfig}
\usepackage{margin}
\usepackage{times}
\fi\fi
\if\papercls \apjcls
\newcommand\aas{\ref@jnl{AAS Meeting Abstracts}}
\newcommand\dps{\ref@jnl{AAS/DPS Meeting Abstracts}}
\newcommand\maps{\ref@jnl{MAPS}}
\else\if\papercls \othercls
\usepackage{astjnlabbrev-jh}
\fi\fi

\if\papercls \aastexcls
\hypersetup{citecolor=blue, 
            linkcolor=blue, 
            menucolor=blue, 
            urlcolor=blue}  
\else
\usepackage[
bookmarks=true,           
bookmarksnumbered=true,   
colorlinks=true,          
citecolor=blue,           
linkcolor=blue,           
menucolor=blue,           
urlcolor=blue,            
linkbordercolor={0 0 1},  
pdfborder={0 0 1},
frenchlinks=true]{hyperref}
\fi

\usepackage{amsmath}
\usepackage{color}
\usepackage{ulem}
\usepackage{bold-extra}
\hypersetup{plainpages=true, colorlinks=true, anchorcolor=blue,
  linkcolor=blue, citecolor=blue, bookmarks=false}

\usepackage{natbib}
\defcitealias{kim17}{Paper I}
\defcitealias{kim18}{Paper II}

\received{May 15, 2019}
\accepted{August 19, 2019}

\newcommand{\Athena}{{\it Athena}\ignorespaces}
\newcommand{\HII}{\ion{H}{2}\ignorespaces}
\newcommand{\p}{\partial}

\newcommand{\pc}{\,{\rm pc}}

\newcommand{\second}{\,{\rm s}}
\newcommand{\yr}{\,{\rm yr}}
\newcommand{\Myr}{\,{\rm Myr}}

\newcommand{\cm}{\,{\rm cm}}
\newcommand{\gram}{\,{\rm g}}
\newcommand{\kms}{\,{\rm km}\,{\rm s}^{-1}}
\newcommand{\Msun}{\,M_{\odot}}
\newcommand{\Lsun}{\,L_{\odot}}
\newcommand{\Kel}{\,{\rm K}}

\newcommand{\eV}{{\,{\rm eV}}}
\newcommand{\Sunit}{\,M_{\odot}\,{\rm pc^{-2}}}

\newcommand{\Mcl}{M_{\rm 0}}
\newcommand{\Rcl}{R_{\rm 0}}
\newcommand{\Sigmacl}{\Sigma_{\rm 0}}
\newcommand{\tffcl}{t_{\rm ff,0}}

\newcommand{\Rh}{R_{\rm h}}

\newcommand{\Qi}{Q_{\rm i}}
\newcommand{\Qphot}{Q_{\rm gas,i}}
\newcommand{\nion}{n_{\rm i}}
\newcommand{\nrms}{n_{\rm i,rms}}
\newcommand{\nH}{n_{\rm H}}
\newcommand{\nHI}{n_{\rm H^0}}

\newcommand{\nelec}{n_e}
\newcommand{\xn}{x_{\rm n}}
\newcommand{\alphaB}{\alpha_{\rm B}}

\newcommand{\fesc}{f_{\rm esc}}

\newcommand{\fesci}{f_{\rm esc,i}}
\newcommand{\fescn}{f_{\rm esc,n}}
\newcommand{\fion}{f_{\rm gas,i}}
\newcommand{\fdusti}{f_{\rm dust,i}}
\newcommand{\fdustn}{f_{\rm dust,n}}

\newcommand{\fescicum}{f_{\rm esc,i}^{\rm cum}}
\newcommand{\fescncum}{f_{\rm esc,n}^{\rm cum}}
\newcommand{\fioncum}{f_{\rm gas,i}^{\rm cum}}
\newcommand{\fdusticum}{f_{\rm dust,i}^{\rm cum}}
\newcommand{\fdustncum}{f_{\rm dust,n}^{\rm cum}}

\newcommand{\kappad}{\kappa_{\rm d}}

\newcommand{\taud}{\tau_{\rm d}}
\newcommand{\taui}{\tau_{\rm i}}
\newcommand{\tauext}{\tau^{\rm ext}}
\newcommand{\tauc}{\tau^{\rm c}}
\newcommand{\taucmean}{\langle\tau^{\rm c}\rangle_{\Omega}}

\newcommand{\muc}{\mu^{\rm c}}
\newcommand{\sigmac}{\sigma^{\rm c}}
\newcommand{\taueff}{\tau_{\rm eff}}
\newcommand{\SFE}{\varepsilon_{*}}

\begin{document}

\title{Modeling UV Radiation Feedback from Massive Stars: \\III. Escape of
  Radiation from Star-forming Giant Molecular Clouds}

\shorttitle{Escape of UV Radiation from GMCs} %
\shortauthors{Kim, Kim, \& Ostriker} %

\author[0000-0001-6228-8634]{Jeong-Gyu Kim}
\affiliation{Department of Astrophysical Sciences, Princeton University,
Princeton, NJ 08544, USA}
\affiliation{Department of Physics \& Astronomy, Seoul National University,
Seoul 08826, Republic of Korea}

\author[0000-0003-4625-229X]{Woong-Tae Kim}
\affiliation{Department of Astrophysical Sciences, Princeton University,
Princeton, NJ 08544, USA}
\affiliation{Department of Physics \& Astronomy, Seoul National University,
Seoul 08826, Republic of Korea}
\affiliation{Center for Theoretical Physics (CTP), Seoul National University, Seoul 08826, Republic of Korea}

\author[0000-0002-0509-9113]{Eve C.~Ostriker}
\affiliation{Department of Astrophysical Sciences, Princeton University,
Princeton, NJ 08544, USA}

\email{kimjg@astro.princeton.edu, wkim@astro.snu.ac.kr, eco@astro.princeton.edu}

\begin{abstract}
  Using a suite of radiation hydrodynamic simulations of star cluster formation
  in turbulent clouds, we study the escape fraction of ionizing (Lyman
  continuum) and non-ionizing (FUV) radiation for a wide range of cloud masses
  and sizes. The escape fraction increases as \HII\ regions evolve and reaches
  unity within a few dynamical times. The cumulative escape fraction before the
  onset of the first supernova explosion is in the range 0.05--0.58; this is
  lower for higher initial cloud surface density, and higher for less massive
  and more compact clouds due to rapid destruction. Once \HII\ regions break out
  of their local environment, both ionizing and non-ionizing photons escape from
  clouds through fully ionized, low-density sightlines. Consequently, dust
  becomes the dominant absorber of ionizing radiation at late times and the
  escape fraction of non-ionizing radiation is only slightly larger than that of
  ionizing radiation. The escape fraction is determined primarily by the mean
  $\langle \tau\rangle$ and width $\sigma$ of the optical-depth distribution in
  the large-scale cloud, increasing for smaller $\langle \tau\rangle$ and/or
  larger $\sigma$. The escape fraction exceeds (sometimes by three orders of
  magnitude) the naive estimate $e^{-\langle \tau\rangle}$ due to non-zero
  $\sigma$ induced by turbulence. We present two simple methods to estimate,
  within $\sim20\%$, the escape fraction of non-ionizing radiation using the
  observed dust optical depth in clouds projected on the plane of sky. We
  discuss implications of our results for observations, including inference of
  star formation rates in individual molecular clouds, and accounting for
  diffuse ionized gas on galactic scales.
\end{abstract}


\section{Introduction}\label{s:intro}

Intense ultraviolet (UV) radiation produced by massive OB stars regulates
heating, ionization, and chemistry in the interstellar medium (ISM), both within
and beyond star-forming clouds. Lyman continuum (LyC) photons capable of
ionizing hydrogen (with energy $h\nu > 13.6 \eV$) create \HII\ regions around
massive stars or clusters in giant molecular clouds (GMCs). Due to elevated
local pressure, \HII\ regions dynamically expand and strongly affect GMC
evolution and star formation within them \citep[][and references
therein]{mck07,kru14,dal15,kru18b}. Far-UV (FUV) photons (with energy
$6.0\eV < h\nu < 13.6 \eV$) can penetrate deep into GMCs to ionize and
dissociate numerous atomic and molecular species, forming photodissociation
regions. Emission lines from these photodissociation regions are crucial probes
of the physical conditions in star-forming GMCs \citep{hol99}.

Some fraction of UV photons emitted by massive stars can escape from GMCs
without being absorbed by gas and dust. The leakage of ionizing photons from
``classical'' \HII\ regions embedded in GMCs is the most likely source of
photoionization of warm ionized gas in the diffuse ISM \citep[the diffuse
ionized gas (DIG) or warm ionized medium (WIM); e.g.,][]{rey84,haf09}. The
further escape of stellar ionizing photons from galaxies into the intergalactic
medium is crucial to the reionization history of the early universe
\citep[e.g.,][]{loe01,rob10,bro11,wis19}. It is estimated that an escape
fraction of least 10--30\% is required for typical stellar populations in
star-forming galaxies to induce significant reionization at a redshift
$7\lesssim z\lesssim 9$ (e.g., \citealt{bou11, fin12, rob15}, but see
\citealt{fin19}), placing demanding requirements on the cloud-scale escape
fraction.

Equally important to the escape of ionizing photons, 
FUV photons escaping into the diffuse ISM
determine the strength of the interstellar background radiation field
\citep{par03}. Via the photoelectric effect on dust, this FUV radiation provides
the dominant form of heating for the diffuse atomic ISM
\citep[e.g][]{wol95,wol03}, amounting to most of the gas mass in galaxies.
Diffuse FUV heating controls the thermal pressure in the diffuse ISM
($P_{\rm th} \propto J_{\rm FUV}$), providing partial support against gravity
and contributing to the self-regulation of star formation on galactic scales
\citep[e.g.,][]{ost10,kimcg13}.


Despite the importance of escaping LyC and FUV radiation from star-forming
regions, direct observational constraints on the escape fraction from GMCs have
been scarce and remain uncertain
\cite[e.g.,][]{smi07,dor13,vog08,pel12,bin18,mcl19}. \citet{smi07} estimated the
escape fraction of ionizing radiation, $\fesci$, from the Carina Nebula, using
spectral classifications of individual massive stars to establish the baseline
for the 
total ionizing photon production rate \citep{smi06}. By comparing to the
observed free-free emission, they estimated that $\sim 25\%$ of ionizing photons
escape through holes in the nebula. They also estimated the escape fraction of
non-ionizing radiation, $\fescn \sim 20\%$, by comparing the total FUV output of
known OB stars with the infrared (IR) emission from the cool dust component.
\citet{dor13} took a similar approach to estimate $\fesci \sim 6\%$ for the 30
Doradus region. \citet{vog08} compared the observed (extinction-corrected)
H$\alpha$ luminosity of \HII\ regions in the Large Magellanic Cloud with the
expected H$\alpha$ luminosity from the observed stellar content, finding that
$\sim 20$--$30\%$ of \HII\ regions are density-bounded. \citet{pel12}
investigated $\fesci$ of individual \HII\ regions in the Large and Small
Magellanic Clouds based on the optical depth of \HII\ regions from the map of
emission-line ratios such as [\ion{S}{2}]/[\ion{O}{3}]. They found the
luminosity-weighted escape fractions amount to $\sim 0.4$, dominated by the most
luminous \HII\ regions.

An additional but more indirect constraint on $\fesci$ is obtained by measuring
the contribution of diffuse H$\alpha$ emission relative to the total (diffuse +
classical \HII\ regions) H$\alpha$ emission in external galaxies. Provided that
photons from massive stars in young clusters dominate in ionizing the diffuse
gas and that the galaxy-scale escape fraction is low, the diffuse H$\alpha$
fraction probes the (globally averaged) cloud-scale escape fraction. Deep
H$\alpha$ images of nearby galaxies show significant ($\sim 20$--$60\%$) diffuse
emission across their disks \citep[e.g.,][]{fer96,hoo96,zur00,oey07,kre16,
  lac18,poe19}. For a sample of 109 \ion{H}{1}-selected nearby galaxies,
\citet{oey07} found that the mean fraction of diffuse H$\alpha$ emission is
$0.59$, with a systematically lower diffuse fraction in starburst galaxies.
\citet{wei18} found that 60\% of the H$\alpha$ emission comes from the diffuse
ionized gas in the central regions of the interacting Antennae galaxy.
\citet{wei18} also estimated $\fesci$ of individual \HII\ regions by comparing
their H$\alpha$ luminosity with the LyC production rate estimated from the
catalog of young star clusters inside \HII\ regions, and found that the overall
cloud-scale escape fraction is consistent with the diffuse fraction.

For a complete accounting, it is necessary to allow for dust absorption of
ionizing radiation, and the H$\alpha$ emission must be extinction-corrected, for
both star-forming regions and diffuse gas. These adjustments can be quite
important, and ``raw'' H$\alpha$ fractions may be misleading subject to the
relative roles of dust in the diffuse and dense ISM. For example, the relative
probability of losing LyC photons to ionization vs. dust absorption depends
inversely on the ionization parameter \citep[e.g.,][]{dop03}, which is higher in
\HII\ regions than the diffuse ISM; H$\alpha$ from dense star forming regions is
strongly extincted compared to H$\alpha$ from the diffuse ISM.

While there are some (albeit uncertain) empirical estimates regarding escape
fractions of photons from star forming regions, on the theory side current
understanding is more limited. Theoretical models of the internal structure of
\HII\ regions are mostly limited to spherical, ionization-bounded \HII\ regions
with $\fesci = 0$ \citep[e.g.,][]{pet72,ino02,dop03,dra11b}, so they are not
useful for studying escape fractions \citep[but see][]{rah17}. Massive stars
form in clusters deeply embedded within dense cores of GMCs \citep{tan14}, so
that nascent \HII\ regions are highly compact and ionization bounded
\citep{hoa07}. However, expansion with evolution leads to a situation where
\HII\ regions become density bounded and exhibit extended envelopes
\citep[e.g.,][]{kim01,kim03}, since turbulence and stellar feedback create
low-density, optically-thin holes through which radiation can escape. As the
processes involved are highly nonlinear, time dependent, and lacking in any
simplifying symmetry, radiation hydrodynamic (RHD) simulations are essential for
quantifying photon escape fractions.

In recent years, several numerical studies have investigated the UV escape
fraction on cloud scales using simulations of star cluster formation with
self-consistent radiation feedback \citep{dal12,dal13,wal12,how17,ras17,kimm19}.
For instance, \citet{dal12,dal13} performed simulations of cloud disruption with
the effects of photoionization feedback included. Using cloud models with the
initial virial parameter of $\alpha_{\rm vir,0}=1.4$ or $4.6$, they found that
$\fesci$ increases with time as clouds are dispersed by feedback. For clouds
with low escape velocities and large virial ratios, $\fesci$ reaches
$\gtrsim 50\%$ before the onset of first supernovae ($3\Myr$ after massive star
formation). \citet{how17,how18} simulated cluster formation in an initially
unbound GMCs with $\alpha_{\rm vir,0}=3$ and masses $10^4$--$10^6\Msun$ under
the influence of both photoionization and radiation pressure feedback. They
studied the temporal changes of $\fesci$ in these models during the first
$\sim 5 \Myr$ of the cloud evolution after massive star formation. They found
that $\fesci$ is highly variable with time because the surrounding gas is highly
turbulent, and that the highest escape fraction ($\fesci > 0.9$) is achieved
only in intermediate cloud masses ($\sim 5\times 10^4\Msun$). More recently,
\citet{kimm19} performed RHD simulations of cloud destruction by the combined
action of photoionization, radiation pressure, and supernovae explosions, also
following evolution of several chemical species. They found a strong positive
relationship between the star formation efficiency and the time-averaged LyC
escape fraction, as stronger feedback clears away the gas and lowers the neutral
gas covering fraction more rapidly.

Although the previous numerical studies mentioned above have greatly improved
our understanding of the cloud-scale escape fraction, they are not without
limitations. One limitation has been in the radiation model and cloud
  parameter space. For example, the simulations of \citet{dal12,dal13} did not
incorporate the effects of dust absorption on $\fesci$. \citet{how17,how18}
considered clouds with fixed mean density, so that they covered only a narrow
range of the parameter space. \citet{kimm19} mostly focused on two
  basic cloud models while considering low and high SFE and low and high
  metallicity. In addition, most of these previous studies focused only on
$\fesci$, but did not study $\fescn$, which is crucial for understanding
emission from star-forming clouds as well as the interstellar radiation field.
\citet{ras17}, on the other hand, studied the escape of non-ionization
radiation, but did not include ionizing photons.

In a series of numerical RHD studies, we have been investigating star cluster
formation in turbulent GMCs with diverse properties, as well as the impact of
stellar radiation feedback on cloud disruption. In \citet[][hereafter Paper
I]{kim17}, we presented the implementation and tests of our numerical RHD
method, which adopts the adaptive ray-tracing algorithm of \citet{abe02} for
point source radiative transfer. In \citet[][hereafter Paper II]{kim18}, we
presented results from models with a range of GMC size and mass, assessing the
dependence of star formation efficiency (SFE) and cloud lifetime on the cloud
surface density, quantifying mass loss due to photoevaporation, and analyzing
momentum injection and disruption driven by gas and radiation pressure forces.

In this paper, we reanalyze the simulations presented in \citetalias{kim18},
focusing on the escape fractions of both ionizing and non-ionizing radiation.
Our main objectives are as follows. First, we explore how $\fesci$ and $\fescn$
from star-forming GMCs vary with time and calculate the cumulative escape
fractions before the epoch of the first supernova. Second, we compare the
fraction of ionizing radiation absorbed by gas and dust with the prediction from
analytic solutions for static, spherical, ionization-bounded \HII\ regions.
Third, we investigate how closely $\fesci$ is related to $\fescn$. Fourth, we
investigate how escape fractions can be estimated from the angular distribution
of the optical depth seen from the sources. Lastly, we propose methods to
estimate the escape fractions from
the mean optical depth or the area distribution of the optical depth projected
along the line of sight of an external observer.

The rest of this paper is organized as follows. In Section~\ref{s:method}, we
briefly describe our numerical methods and initial conditions of the
simulations. In Section~\ref{s:evol}, we present results on the overall
evolution of the simulated clouds. This includes quantifying the fractions of
photons that are absorbed by gas, by dust, and that escape from the clouds. In
Section~\ref{s:PDF}, we calculate optical depth distributions as seen from the
luminosity center or by an external observer, and we relate these distributions
to the measured escape fractions. In Section~\ref{s:summary}, we summarize and
discuss our main results. In Appendix~\ref{s:appendixA}, we develop and apply a
subgrid model to explore how radiation absorbed in the immediate vicinity of
sources (which we do not numerically resolve) may affect SFE estimates.
In Appendix~\ref{s:appendixB}, we explore the potential effect of dust
  destruction in ionized gas on the escape fraction of radiation.

\section{Numerical Methods}\label{s:method}

We study the escape fractions of ionizing and non-ionizing radiation from 
star-forming, turbulent GMCs based on a suite of RHD simulations
presented in \citetalias{kim18}. These simulations were performed using the
grid-based magnetohydrodynamics code \Athena\ \citep{sto08}, equipped with
modules for self-gravity, sink particles, and point source radiative transfer.
In this section, we briefly summarize the numerical methods and cloud models.
The reader is referred to \citetalias{kim17} and \citetalias{kim18} for
technical details as well as more quantitative results.

\subsection{Radiation Hydrodynamics Scheme}

We solve the equations of hydrodynamics in conservation form using the van Leer
type time integrator \citep{sto09}, HLLC Riemann solver, and piecewise linear
spatial reconstruction method. We employ the sink particle method of
\citet{gon13} to handle cluster formation and ensuing mass accretion. A
Lagrangian sink particle (representing a subcluster of young stars) is created
if a gas cell (1) has the density above a  threshold value set by the
Larson-Penston self-gravitating collapse solution imposed at the grid scale,
(2) has a converging
velocity field around
it, and (3) is at the local minimum of the gravitational potential. The gas mass
that is accreted onto a sink particle is calculated based on the fluxes returned
by the Riemann solver at the boundary faces of a $3^3$-cell control volume
surrounding it. The gravitational potential from gas and stars is computed using
the fast Fourier transform Poisson solver with the vacuum boundary conditions 
\citep{ski15}.

The UV radiative output of a star cluster is calculated based on the
mass-luminosity relation obtained from Monte-Carlo simulations for the spectra
of a zero-age main sequence population with a Chabrier initial mass function
(IMF) \citep{kim16}. For a given total cluster mass $M_{*,{\rm tot}}$, we
compute the total UV luminosity
$L = L_{\rm i} + L_{\rm n} \equiv \Psi M_{*,{\rm tot}}$ and the total ionizing
photon rate $\Qi = L_{\rm i}/(h\nu_{\rm i}) \equiv \Xi M_{*,{\rm tot}}$, where
$L_{\rm i}$ and $L_{\rm n}$ refer to the luminosity of ionizing and non-ionizing
radiation, respectively, and $h\nu_{\rm i} = 18 \eV$ is the mean energy of
ionizing photons. The light-to-mass ratios $\Psi$ and $\Xi$ are in general
functions of $M_{*,{\rm tot}}$. To allow for the effects of incomplete sampling
of the IMF at the high-mass end, we fit $\Psi$ and $\Xi$ to the median values of
multiple realizations of the IMF. It turns out that
$\Psi \rightarrow 912 \Lsun \Msun^{-1}$ and
$\Xi \rightarrow 5.05 \times 10^{46} \second^{-1} \Msun^{-1}$ in the limit of a
fully sampled IMF ($M_{*,{\rm tot}} \gtrsim 10^4 \Msun$), while they sharply
decline with decreasing $M_{*,{\rm tot}} \lesssim 10^3 \Msun$. We treat the
instantaneous set of star particles as a single cluster to determine the total
luminosity, and the luminosity of each sink particle is assigned in proportion
to its mass. We do not consider temporal evolution of $\Psi$ and $\Xi$ in the
present work.

We adopt the adaptive ray-tracing method \citep{abe02} to track the
radiation field emitted from multiple sources. Photon packets injected at the
position of each source particle propagate along the rays whose directions are
determined by the HEALPix scheme of \citet{gor05}, which divides the unit
sphere into equal-area pixels. Rays are split adaptively to ensure that each
cell is crossed by at least four rays per source. The length of a line segment
passing through the cell is used to calculate the optical depth that is required
to evaluate the volume-averaged radiation energy densities
$\mathcal{E}_{\rm i}, \mathcal{E}_{\rm n}$ and fluxes
$\mathbf{F}_{\rm i}, \mathbf{F}_{\rm n}$ in the ionizing and non-ionizing
frequency bins at every cell.

As sources of UV opacity, we consider absorption of ionizing photons by neutral
hydrogen and that of both ionizing and non-ionizing photons by dust. We adopt
constant values of
$\sigma_{\rm ph} = 6.3 \times 10^{-18}\,{\rm cm}^2\,{\rm H}^{-1}$ for the
photoionization cross section \citep{kru07}\footnote{The adopted cross section
  is the value at the Lyman edge ($h\nu = 13.6\eV$). We have verified that the
  use of a more realistic photoionization cross section averaged over the
  stellar spectrum (a factor of $\sim 2$ smaller) increases the neutral fraction
  within the primarily-ionized regions (see Equation \eqref{eq:xneq1}), but does
  not affect other simulation outcomes.} and
$\sigma_{\rm d,i/n} = \sigma_{\rm d} = 1.17 \times 10^{-21}\,{\rm cm}^2\,{\rm
  H}^{-1}$ for the dust absorption cross section per hydrogen (or cross section
per unit gas mass
$\kappad = \sigma_{\rm d}/\mu_{\rm H} = 500 \cm^2\,\gram^{-1}$, with
$\mu_{\rm H}=1.4 m_{\rm H}$ being the mean molecular weight)
\citep{dra11b}.\footnote{ We ignore dust scattering altogether. It should be
  noted that in dust models for the diffuse ISM, the scattering is strongest in
  the forward direction with albedo $\sim 0.2$--$0.4$ in the UV wavelengths, so
  that the neglect of scattering may be a reasonably good approximation
  \citep[e.g.,][]{gla19}.} We discuss the potential impact of dust destruction
in ionized regions on the escape fraction in Section~\ref{s:ddest}. The
resulting radiation energy and flux densities are used to calculate the local
photoionization rate
$\mathcal{I} = \nHI \sigma_{\rm ph} c\mathcal{E}_{\rm i}/(h\nu_{\rm i})$ and
radiation pressure force
$\frac{n_{{\rm H^0}}\sigma_{\rm ph}}{c} \mathbf{F}_{\rm i} + \frac{n_{\rm H}
  \sigma_{\rm d}}{c}(\mathbf{F}_{\rm i} + \mathbf{F}_{\rm n})$ on dusty gas,
where $n_{\rm H}$ and $n_{\rm H^0}$ are the number density of total and neutral
hydrogen, respectively.

We solve the continuity equation for neutral hydrogen including source and sink
terms due to recombination and photoionization, adopting case B recombination
rate coefficient
$\alphaB = 3.03 \times 10^{-13}\,{\rm cm}^{3}\,{\rm s}^{-1}(T/8000\Kel)^{-0.7}$
\citep{kru07}. The source and sink terms are explicitly updated every substep in
an operator split fashion. The gas temperature is set to vary smoothly as a
function of the neutral gas fraction between $20 \Kel$ and $8000 \Kel$,
corresponding to the temperature of fully neutral and fully ionized gas,
respectively. The use of constant equilibrium temperature is a good
approximation if the gas cooling time is short compared to the dynamical
timescale \citep[e.g.,][]{lef94}. Although our model cannot represent the
detailed thermal structure of \HII\ regions because we neglect ionization of
helium and do not follow specific heating/cooling processes, it still captures
the essential physics needed to follow the dynamics of \HII\ regions with
self-consistent star formation, which is crucial for modeling escape of
radiation.

\subsection{Initial and Boundary Conditions}

We establish initial conditions of our model clouds following \citet{ski15}. We
start with a uniform-density gas sphere with mass $\Mcl$ and radius $\Rcl$
placed at the center of a computational box, surrounded by a tenuous medium with
density $10^3$ times lower than the cloud. The box is a cube with each side
$L_{\rm box} = 4\Rcl$. Our standard resolution is $N=256$ cells in one
direction, although we also run simulations with $N=128$ or $512$ to test
convergence for the fiducial model. Initially, the cloud is completely neutral
and seeded by a (decaying) turbulent velocity field with a power
spectrum $|\delta \mathbf{v}_k|^2 \propto k^{-4}$ over the wavenumber
  range $k \in \left[2,64\right] \times 2\pi/L_{\rm box}$. The initial cloud is
  set to be marginally bound, with the initial virial parameter
  $\alpha_{\rm vir,0} \equiv 5 \sigma_{v,0}^2 \Rcl/(3G\Mcl) = 2$, where
  $\sigma_{v,0}$
  is the turbulent velocity dispersion \citep[e.g.,][]{ber92}.

We adopt strict outflow (diode-like) boundary conditions both at the outer
boundaries of the computational domain and at the boundary faces of the control
volume surrounding each sink particle. The $3^3$ control-volume cells serve as
internal ghost zones within the simulation domain. Because of the presence of a
point mass which is also a source of radiation, gravity and hydrodynamic
variables are unresolved within control volumes, and we do not attempt to model
photon-gas interactions. Instead, we simply allow all of the photons emitted by
a sink particle to emerge from the control volume without absorption,
corresponding to $f_{\rm esc,*}=1$, where $f_{\rm esc,*}$ denotes the escape
fraction of radiation from the control volume (i.e., ``subgrid'' scale). In
reality, photon-gas interactions inside the control volume would lower
$f_{\rm esc,*}$ below unity and thus reduce the overall efficiency of radiation
feedback on cloud scales. In Appendix~\ref{s:appendixA}, we explore the effect of
varying $f_{\rm esc,*}$ on the cloud-scale SFE, using simulations with only
non-ionizing radiation. There, we discuss the plausible range of
$f_{\rm esc,*}$, and show that the final stellar mass in our fiducial
cloud is increased only modestly ($\sim 0.1\Mcl$) if $f_{\rm esc,*}$ is allowed
to drop below unity.

\begin{deluxetable*}{ccccccccccccc}
\tabletypesize{\footnotesize}
\tablewidth{0pt}\tablecaption{Model parameters and simulation results}
\tablehead{
\colhead{Model} &
\colhead{$M_0$} &
\colhead{$R_0$} &
\colhead{$\Sigma_0$} &
\colhead{$n_{\rm H,0}$} &
\colhead{$t_{\rm ff,0}$} &
\colhead{$t_{\rm dest}$} &
\colhead{$Q_{\rm i,max}$} &
\colhead{$f_{\rm esc,i}^{\rm cum}$} &
\colhead{$f_{\rm phot,i}^{\rm cum}$} &
\colhead{$f_{\rm dust,i}^{\rm cum}$} &
\colhead{$f_{\rm esc,n}^{\rm cum}$} &
\colhead{$f_{\rm dust,n}^{\rm cum}$} \\
\colhead{(1)} &
\colhead{(2)} &
\colhead{(3)} &
\colhead{(4)} &
\colhead{(5)} &
\colhead{(6)} &
\colhead{(7)} &
\colhead{(8)} &
\colhead{(9)} &
\colhead{(10)} &
\colhead{(11)} &
\colhead{(12)} &
\colhead{(13)}}
\startdata
{\tt M1E5R50} & $10^5$ & $50.0$ & $12.7$ & $ 5.5$ & $18.5$ & $10.4$& $10^{50.2}$ & $0.58$ & $0.28$ & $0.14$ & $0.72$ & $0.28$ \\ 
{\tt M1E5R40} & $10^5$ & $40.0$ & $19.9$ & $10.8$ & $13.2$ & $10.4$& $10^{50.4}$ & $0.48$ & $0.34$ & $0.18$ & $0.61$ & $0.39$ \\ 
{\tt M1E5R30} & $10^5$ & $30.0$ & $35.4$ & $25.5$ & $ 8.6$ & $ 8.2$& $10^{50.6}$ & $0.45$ & $0.36$ & $0.20$ & $0.56$ & $0.44$ \\ 
{\tt M1E4R08} & $10^4$ & $ 8.0$ & $49.7$ & $134.7$ & $ 3.7$ & $ 3.9$& $10^{49.0}$ & $0.47$ & $0.38$ & $0.15$ & $0.58$ & $0.42$ \\ 
{\tt M1E6R80} & $10^6$ & $80.0$ & $49.7$ & $13.5$ & $11.8$ & $ 8.2$& $10^{51.7}$ & $0.12$ & $0.56$ & $0.33$ & $0.23$ & $0.77$ \\ 
{\tt M5E4R15} & $5 \times 10^4$ & $15.0$ & $70.7$ & $102.2$ & $ 4.3$ & $ 6.3$& $10^{50.5}$ & $0.37$ & $0.39$ & $0.24$ & $0.46$ & $0.54$ \\ 
{\tt \textbf{M1E5R20}} & $\bf 10^5$ & $\bf 20.0$ & $\bf 79.6$ & $\bf 86.2$ & $\bf  4.7$ & $\bf  6.5$& $10^{50.8}$ & $\bf 0.30$ & $\bf 0.38$ & $\bf 0.32$ & $\bf 0.39$ & $\bf 0.61$ \\ 
{\tt M1E4R05} & $10^4$ & $ 5.0$ & $127.3$ & $551.8$ & $ 1.9$ & $ 3.8$& $10^{49.7}$ & $0.35$ & $0.40$ & $0.26$ & $0.42$ & $0.58$ \\ 
{\tt M1E6R45} & $10^6$ & $45.0$ & $157.2$ & $75.7$ & $ 5.0$ & $ 6.9$& $10^{52.0}$ & $0.10$ & $0.54$ & $0.36$ & $0.14$ & $0.86$ \\ 
{\tt M1E5R10} & $10^5$ & $10.0$ & $318.3$ & $689.7$ & $ 1.7$ & $ 3.7$& $10^{51.1}$ & $0.12$ & $0.50$ & $0.38$ & $0.16$ & $0.84$ \\ 
{\tt M1E4R03} & $10^4$ & $ 3.0$ & $353.7$ & $2554.6$ & $ 0.9$ & $ 2.7$& $10^{49.9}$ & $0.49$ & $0.28$ & $0.23$ & $0.52$ & $0.48$ \\ 
{\tt M1E6R25} & $10^6$ & $25.0$ & $509.3$ & $441.4$ & $ 2.1$ & $ 5.0$& $10^{52.3}$ & $0.05$ & $0.62$ & $0.32$ & $0.07$ & $0.93$ \\ 
{\tt M1E4R02} & $10^4$ & $ 2.0$ & $795.8$ & $8621.6$ & $ 0.5$ & $ 1.9$& $10^{50.3}$ & $0.45$ & $0.38$ & $0.17$ & $0.47$ & $0.53$ \\ 
{\tt M1E5R05} & $10^5$ & $ 5.0$ & $1273.2$ & $5517.8$ & $ 0.6$ & $ 2.1$& $10^{51.4}$ & $0.38$ & $0.38$ & $0.24$ & $0.39$ & $0.61$ \\ 
{\tt M1E5R20\_N128} & $10^5$ & $20.0$ & $79.6$ & $86.2$ & $ 4.7$ & $ 5.3$& $10^{50.9}$ & $0.25$ & $0.40$ & $0.35$ & $0.35$ & $0.65$ \\ 
{\tt M1E5R20\_N512} & $10^5$ & $20.0$ & $79.6$ & $86.2$ & $ 4.7$ & $ 7.2$& $10^{50.8}$ & $0.28$ & $0.44$ & $0.28$ & $0.36$ & $0.64$ \\ 
\enddata
\tablecomments{Column 1: model name indicating initial cloud mass and radius. Column 2: initial gas mass ($M_\odot$). Column 3: initial radius (${\rm pc}$). Column 4: initial gas surface density ($M_\odot\,{\rm pc}^{-2}$). Column 5: initial number density of H (${\rm cm}^{-3}$). Column 6: initial free-fall time (${\rm Myr}$). Column 7: cloud destruction timescale (${\rm Myr}$). Column 8: maximum ionizing photon production rate (${\rm s}^{-1}$). Column 9: cumulative escape fraction of ionizing photons at $t_{*,0} + 3\Myr$. Column 10: cumulative hydrogen absorption fraction of ionizing photons at $t_{*,0} + 3\Myr$. Column 11: cumulative dust absorption fraction of ionizing photons at $t_{*,0} + 3\Myr$. Column 12: cumulative dust absorption fraction of non-ionizing photons at $t_{*,0} + 3\Myr$.  The fiducial model {\tt M1E5R20} is shown in bold.\label{t:result}}
\end{deluxetable*}

\subsection{Cloud Model}


We consider 14 models that span two orders of magnitude in mass
($10^4\Msun < \Mcl < 10^6\Msun$) and surface density
($12.7\Sunit < \Sigmacl < 1.27 \times 10^3\Sunit$) to explore a range of
star-forming environments. For example, low surface density
($\Sigmacl \sim 10^{2} \Sunit$) and massive ($\Mcl \sim 10^{5}$--$10^{6} \Msun$)
clouds are representative of typical GMCs in the Milky Way and normal spiral
galaxies, whereas high surface density ($\Sigmacl \gtrsim 500 \Sunit$) and low
mass ($\Mcl \lesssim 10^5 \Msun$) clouds correspond to individual
cluster-forming clumps within GMCs \citep[e.g.,][]{tan14}. \footnote{These
  clouds are optically thick to UV radiation
  ($\Sigmacl \gtrsim \kappa_{\rm d,UV}^{-1} \sim 10 \Sunit$) but optically thin
  to dust-reprocessed IR radiation
  ($\Sigmacl \lesssim \kappa_{\rm d,IR}^{-1} \sim 10^3 \Sunit$). The pressure
  from trapped IR radiation is likely to play a dominant role only for clouds in
  extremely high-surface density environments \citep[e.g.,][]{ski15,tsa18}.} The
Columns 1--6 of Table \ref{t:result} respectively list the model names, mass
$\Mcl$, radius $\Rcl$, surface density $\Sigmacl = \Mcl/(\pi \Rcl^2)$, number
density of hydrogen $n_{\rm H,0}$, and free-fall time
$t_{\rm ff,0} = \tfrac{\pi}{2}\sqrt{\Rcl^3/(2G\Mcl)}$ of the initial model
clouds. Given the initial virial parameter $\alpha_{\rm vir,0}=2$ in all our
simulations, the initial turbulent Mach number varies from
$\mathcal{M}_0 = \sigma_{v,0}/c_{\rm s}=$ 6 to 33 for the sound speed of neutral
gas $c_{\rm s}=0.26\kms$. We take the ``Orion-like'' model {\tt M1E5R20} with
$\Mcl = 10^5 \Msun$ and $\Rcl = 20\pc$ as our fiducial case. Models {\tt
  M1E5R20\_N128} and {\tt M1E5R20\_N512} correspond to the fiducial cloud at
different resolution with $N=128$ and 512, respectively.

\subsection{Absorption and Escape Fractions of Radiation}

The escape of ionizing radiation is hindered by absorption
by dust and neutral hydrogen. Let $\Qphot = \int \mathcal{I} dV$ denote the
total photoionization rate and
$Q_{\rm dust,i} = \int \nH \sigma_{\rm d} c\mathcal{E}_{\rm i}/(h\nu_{\rm i})
dV$ the total dust absorption rate. Then, the rate of ionizing photons escaping
from the computational domain is given by
$Q_{\rm esc,i} = \Qi - \Qphot - Q_{\rm dust,i}$. The adaptive ray tracing
calculates $\mathcal{E}_{\rm i}$ at every cell and keeps track of
$Q_{\rm esc,i}$ explicitly, allowing us to calculate the hydrogen absorption
fraction, dust absorption fraction, and escape fraction defined as
\begin{align}
  \fion   &\equiv \frac{\Qphot}{\Qi}, \\
  \fdusti &\equiv \frac{Q_{\rm dust,i}}{\Qi}, \\
  \fesci  &\equiv \frac{Q_{\rm esc,i}}{\Qi}=1-\fion-\fdusti,
\end{align}
respectively. These instantaneous quantities are luminosity-weighted averages
over individual sources. We also calculate the cumulative escape fraction
defined as
\begin{equation}
  f_{\rm esc,i}^{\rm cum} (t') \equiv \dfrac{\int_{t_{*,0}}^{t}Q_{\rm
      i,esc}\,dt}{\int_{t_{*,0}}^{t}
    \Qi\,dt},
\end{equation}
and similarly for the cumulative absorption fractions, $\fioncum$ and
$\fdusticum$. Here, $t_{*,0}$ is the time at which the first sink particle is
created and radiative feedback is turned on, and $t'=t-t_{*,0}$.
We similarly  monitor the dust
absorption fraction $\fdustn$ and the escape fraction $\fescn=1-\fdustn$ of
non-ionizing photons.

\section{Time Evolution}\label{s:evol}

\begin{figure*}[!t]
  \epsscale{1.2}\plotone{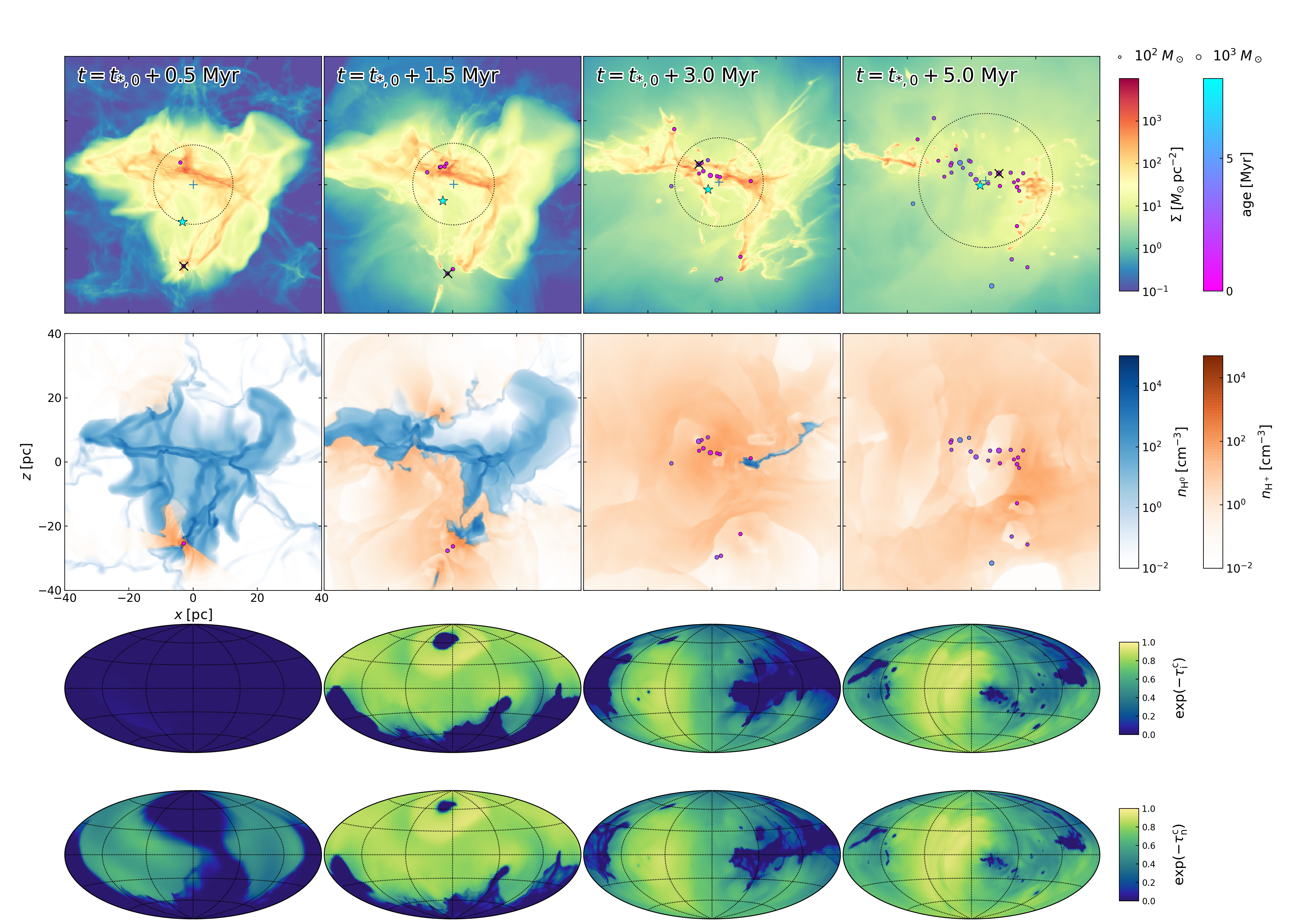}
  \caption{Snapshots of the fiducial model {\tt M1E5R20} ($\Mcl = 10^5 \Msun$
    and $\Rcl = 20 \pc$) at $0.5$, $1.5$, $3.0$, and $5.0\Myr$ (left to right)
    after the first star formation. (top row) Gas surface density projected
    along the $y$-direction. In each panel, the projected positions of star
    particles are indicated by small circles, with age indicated by color. The
    star particle center of mass is indicated with a star symbol. The plus signs
    and the dotted circles mark the center of mass and the half-mass radius of
    the gas in projection, respectively. (middle row) Slices through the most
    massive star particle (marked with $\times$ symbol) of the number density
    of neutral (blue) and ionized (orange) hydrogen in the $x$--$z$ plane. Only
    the star particles within $\Delta y = \pm \Rcl/2$ of the slice are shown.
    (bottom row) Hammer projection maps of the angular distributions of the
    escape probabilities $\exp(-\tau^{\rm c}_{\rm i})$ and
    $\exp(-\tau^{\rm c}_{\rm n})$ for ionizing and non-ionizing radiation as
    seen from the position of the most massive sink particle. At the times shown
    (left to right), the instantaneous escape fraction is
    $\fesci = (15, 31, 37, 59)\%$ for ionizing radiation and
    $\fescn = (31, 39, 45, 65)\%$ for non-ionizing radiation.
  }\label{f:snapshot1}
\end{figure*}

\begin{figure*}[!t]
  \epsscale{1.2}\plotone{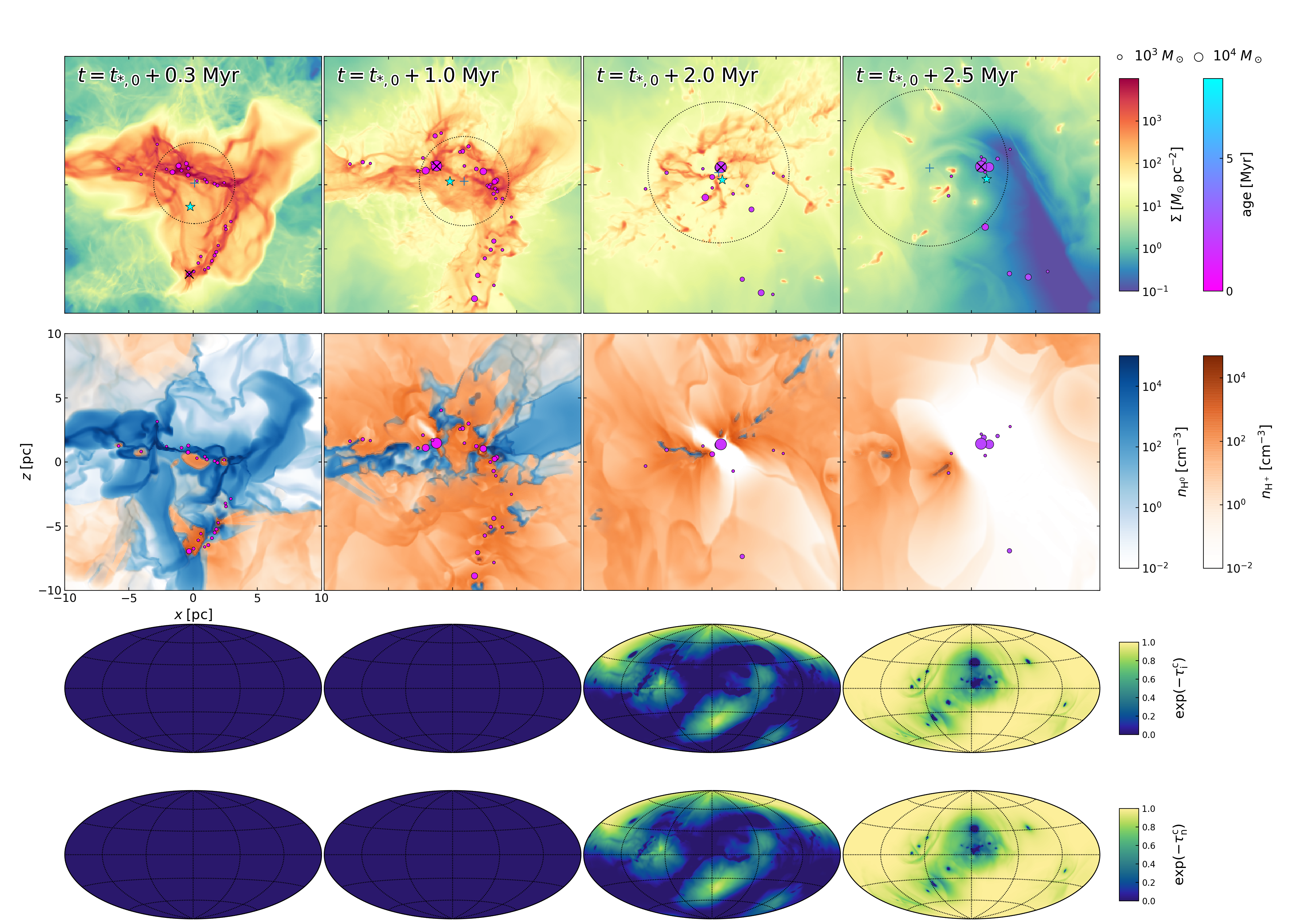}
  \caption{Same as Figure~\ref{f:snapshot1} but for model {\tt M1E5R05} with
    $\Mcl = 10^5\Msun$ and $\Rcl = 5\pc$ at times $t'=$ $0.3$, $1.0$, $2.0$, and
    $2.5\Myr$ after the first star formation. From left to right, the
    instantaneous escape fraction radiation is $\fesci = (2.3, 3.1, 26, 88)\%$
    for ionizing radiation and $\fescn = (3.8, 4.5, 28, 88)\%$ for non-ionizing
    radiation.}\label{f:snapshot2}
\end{figure*}

\begin{figure*}[!t]
  \epsscale{1.2}\plotone{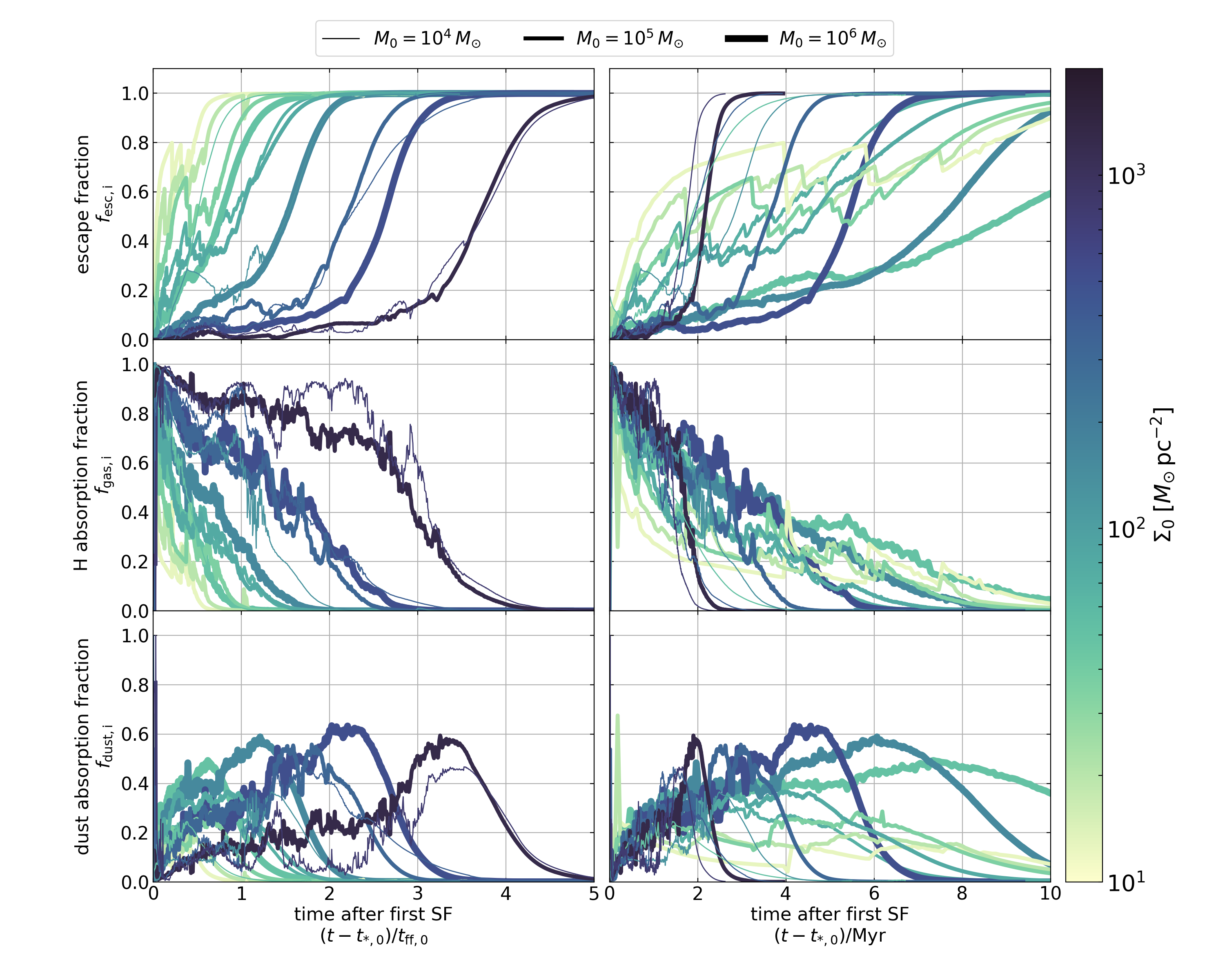}
  \caption{Evolution of the instantaneous escape fraction $\fesci$ (top), the
    hydrogen absorption fraction $\fion$ (middle), and the dust absorption
    fraction $\fdusti$ (bottom) for ionizing radiation.  Time
    is measured from the creation of the first star particle ($t'=t-t_{*,0}$), in
    units of $\tffcl$ (left) or Myr (right). All models are shown, with the
    thickness and color of each
    line indicating the initial cloud mass $\Mcl$ and surface density
    $\Sigmacl$,
    respectively. With small $\tffcl$, high-$\Sigmacl$ clouds evolve
    rapidly: most gas is cleared away and $\fesci$ reaches unity in a few Myr,
    before massive stars explode as supernovae.}\label{f:fevol}
\end{figure*}

We begin by presenting temporal evolution of our fiducial model, with a focus on
the absorption and escape fractions of radiation. Figure~\ref{f:snapshot1}
displays snapshots of the fiducial model ($\Mcl = 10^5 \Msun$ and
$\Rcl = 20 \pc$, with $\Sigmacl = 80\Sunit$ and $n_{\rm H,0}=86 \cm^{-3}$) at
times $t'=0.5$, $1.5$, $3$, and $5 \Myr$, from left to right, after the first
star formation event occurring at $t_{*,0} = 1.87 \Myr$. From top to bottom, the
rows show gas surface density projected along the $y$-axis, slices of the
neutral (blue) and ionized (orange) gas density through the most massive sink
particle in the $x$--$z$ plane, and the Hammer projections of the angular
distributions of the escape probabilities, $\exp(-\tauc_{\rm i})$ and
$\exp(-\tauc_{\rm n})$, of the ionizing and non-ionizing radiation,
respectively, as seen from the most massive sink particle. Here, the
superscripts ``c'' indicate the optical depth calculated outward from a point
within the cloud to the edge of the simulation domain. In the top row, the
dotted circles draw the projected regions enclosing half the total gas mass in
the simulation domain, while the star symbols mark the projected center of mass
of the star particles represented by small circles in the top and second rows.

Cloud evolution is initially driven by supersonic turbulence that readily
produces shock-compressed filaments and clumps. The densest parts of these
structures become gravitationally unstable and soon spawn sink particles.
Ensuing radiation feedback from the sink particles form small \HII\ regions
around them. The \HII\ regions expand outward and break out of the natal clumps,
eventually merging with each other. In this process, the low density gas becomes
rather quickly ionized by the passage of R-type ionization fronts, increasing
its volume fraction from 27\% at $t'=0.2 \Myr$ to 78\% at $t'=1.5 \Myr$ in the
fiducial model. The gas that acquires sufficient radial momentum via thermal and
radiation pressures leaves the simulation domain, which in turn destroys the
cloud and limits the SFE. We measure the cloud destruction timescale as
  the time taken to photoevaporate and/or eject 95\% of the initial cloud mass
  after the onset of radiation feedback (so that only 5\% of the initial cloud
  mass is left over as the neutral phase in the simulation domain), i.e.,
  $t_{\rm dest} \equiv t_{{\rm neu},5\%} - t_{*,0}$, and the net SFE as the
  fraction of the initial cloud mass that turned into stars over the cloud
  lifetime, i.e., $\SFE \equiv M_{*,{\rm final}}/\Mcl$\footnote{It is important
  to stress that the net SFE is a quantity based on the original gas mass and
  final stellar mass, which cannot be directly measured for individual molecular
  clouds; the observed ``instantaneous'' SFE is based on the gas mass and
  stellar mass at the current epoch.}. For the fiducial model, we find
  $t_{\rm dest} = 6.5 \Myr = 1.39 \tffcl$ and
  $\SFE \equiv M_{*,{\rm final}}/\Mcl = 0.13$. As discussed in
  \citetalias{kim18}, the dominant feedback mechanism is photoionization rather
  than radiation pressure: 81\% of the initial cloud mass is lost by
  photoevaporation; the radial momentum injected by thermal pressure in this
  model is $\sim 5$ times higher than that from radiation pressure.

As the Hammer projections in Figure~\ref{f:snapshot1} show, an appreciable
fraction of radiation can escape from the \HII\ regions even before the complete
destruction of the natal clumps. For instance, the instantaneous escape
fractions amount to $\fesci=15\%$ and $\fescn=31\%$ at $t'=0.5 \Myr$
when about $98\%$ of the gas mass is in the neutral phase. This is because
turbulence naturally creates sightlines with low optical depth along which the
\HII\ regions are density bounded, permitting easy escape of radiation. As
star formation continues and gas photoevaporates,
the fraction of solid angle with optically thick,
ionization-bounded sightlines steadily decreases.
This lowers the hydrogen absorption fraction, while increasing
the escape fraction more or less monotonically with time (see Figure 2 of
\citetalias{kim18}). The dust absorption fraction reaches $\fdusti \sim37\%$ at
$t'= 1 \Myr$ and is then maintained at $\sim 30$--$35\%$ for about
$3.5\Myr$ before starting to decline gradually.

While the overall dynamical evolution of other models is qualitatively similar,
we find that the evolution of the absorption and escape fractions depend on the
initial surface density. Figure~\ref{f:snapshot2} plots snapshots of gas surface
density, slices of neutral and ionized volume density, and angular distributions
of the radiation escape probabilities for model {\tt M1E5R05}
($\Mcl = 10^5 \Mcl, \Rcl = 5 \pc$, $\Sigmacl=1.3\times 10^3\Sunit$,
and $n_0=5.5\times 10^3\rm \cm^{-3}$).
Compared to the fiducial run with $\Sigmacl = 80 \Sunit$, the denser
recombination layers and deeper gravitational potential in model {\tt M1E5R05}
make radiation feedback less effective in photoevaporating the neutral gas and
ejecting gas by radiation and thermal pressures, yielding a higher SFE of
$\SFE = 0.51$ (\citetalias{kim18}; see also \citealt{gee17,gru18}). The cloud
destruction time is only $t_{\rm dest} = 2.1\Myr$ since all dynamical processes
are rapid at high density; the free-fall time for this model is just
$\tffcl = 0.6\Myr$. Due to high dust column, trapped \HII\ regions barely break
out and both $\fesci$ and $\fescn$ remain very small during most of the cloud
evolution, as evidenced by the angular distributions of the escape probabilities
shown in Figure~\ref{f:snapshot2}. For example, at $t'= 1.0\Myr$, $\fesci=0.02$
even though the ionized gas fills $\sim 80\%$ of the entire volume. At
$t'=2\Myr$ when star formation is completed and the ionized-gas volume filling
factor is 97\%, $\fesci$ and $\fescn$ increase only to 0.26 and 0.28,
respectively. The cumulative escape and absorption fractions at $t'=2\Myr$ are
$\fescicum = 0.08$, $\fioncum = 0.63$, $\fdusticum = 0.29$, $\fescncum = 0.10$,
and $\fdustncum = 0.90$ in this model.

Figure~\ref{f:fevol} plots the ionizing radiation history of $\fesci$ (top),
$\fion$ (middle), and $\fdusti$ (bottom) as functions of the time for all
models. The line thickness and color indicate $\Mcl$ and $\Sigmacl$,
respectively. For all models, time is measured since the first star formation,
and shown in units of $\tffcl$ and Myr in the left and right panels,
respectively. Overall, $\fesci$ increases as \HII\ regions evolve\footnote{The
  precipitous drops in $\fesci$ (or jumps in $\fion$) in low-$\Sigmacl$ clouds
  occur due to the birth of deeply embedded cluster particles.}, consistent with
expectations and with results from previous simulations
\citep{wal12,dal13,kimm19}. The escape of ionizing radiation is limited
primarily by photoionization in early evolutionary stages and by dust absorption
in late stages. The dust absorption fraction peaks slightly before cloud
destruction, and vanishes as the remaining gas is cleared out. Although
higher-$\Sigmacl$ clouds appear to live longer in terms of $\tffcl$, they are
actually destroyed earlier in real time. Note that clouds with
$\Sigmacl \gtrsim 300 \Sunit$ and $\Mcl \le 10^5 \Msun$ ({\tt M1E4R03}, {\tt
  M1E4R02}, {\tt M1E5R05}) are destroyed in less than $3\Myr$ after the onset of
star formation (Column 7 in Table~\ref{t:result}), resulting in substantial
escape of radiation before the advent of supernova explosions
(Section~\ref{s:fesccum}).

\subsection{Comparison with Spherical Models}

\begin{figure}[!t]
  \epsscale{1.15}\plotone{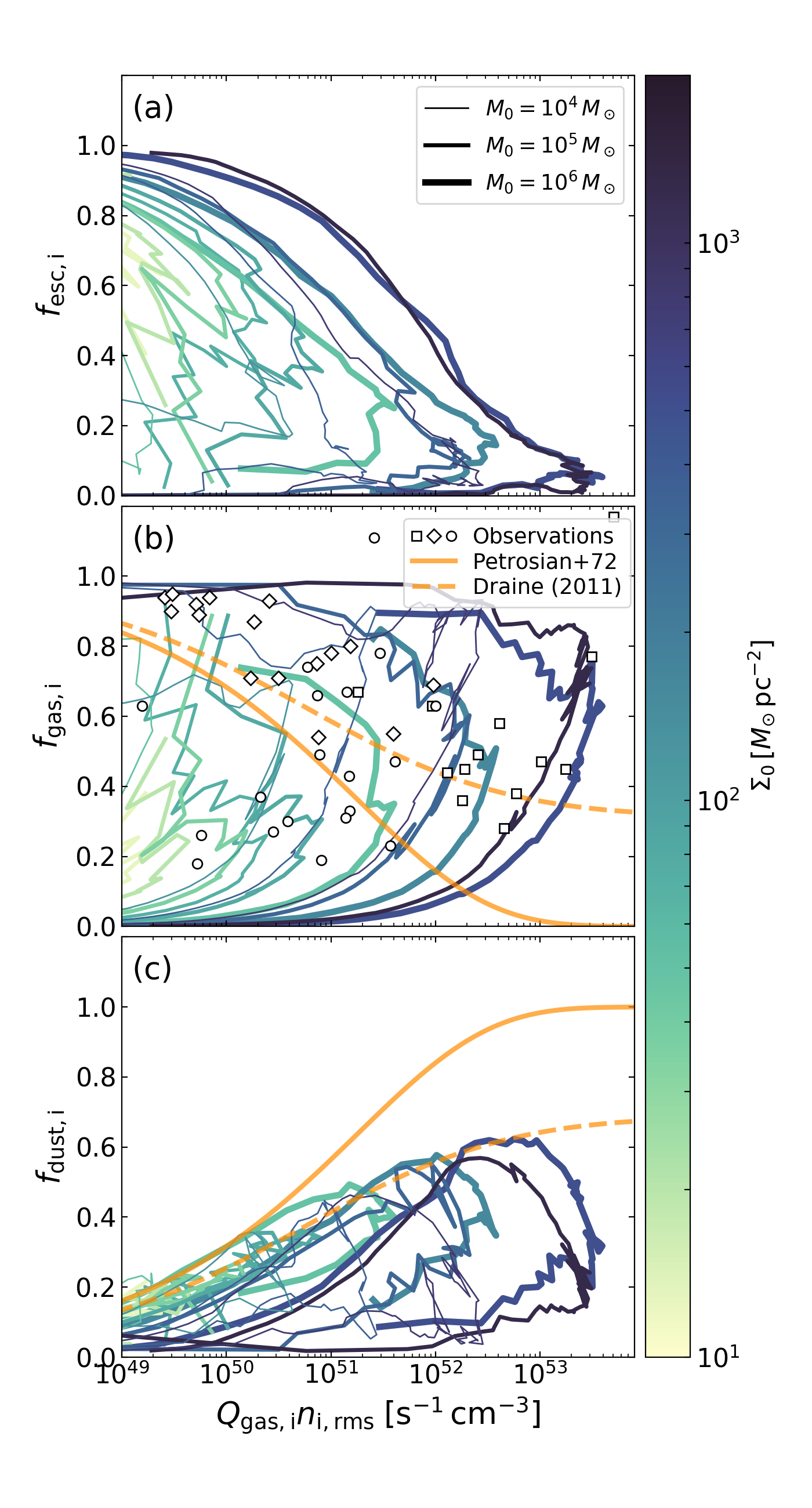}
  \caption{Dependence of (a) the instantaneous escape fraction $\fesci$, (b) the
    hydrogen absorption fraction $\fion$, and (c) the dust absorption fraction
    $\fdusti$ of ionizing radiation on the product of the total photoionization
    rate $\Qphot$ and the rms number density of the ionized gas $\nrms$.
    The thickness and color of each line indicate the initial gas mass $\Mcl$
    and surface density $\Sigmacl$ in the model. 
    The orange dashed and solid lines in (b) and (c) draw the predictions of the
    theoretical models for embedded ($\fesci = 0$), spherical, static \HII\
    regions with \citep{dra11b} and without \citep{pet72} the effects of
    radiation pressure, respectively.
    The open symbols in (b) are the observational estimates for $\fion$ from the
    Galactic \HII\ regions by \citeauthor{ino01} (\citeyear{ino01}, diamonds),
    \citeauthor{ino02} (\citeyear{ino02}, squares), and \citeauthor{bin18}
    (\citeyear{bin18}, circles).}\label{f:Qeffn}
\end{figure}

It is interesting to compare the hydrogen and dust absorption fractions
calculated in our simulations with the analytic predictions for static,
spherical, ionization-bounded ($\fesci=0$) \HII\ regions. For this purpose,
Figure~\ref{f:Qeffn} plots as various lines (a) the escape fraction ($\fesci$),
(b) the hydrogen absorption fraction ($\fion$), and (c) the dust absorption
fraction ($\fdusti$) of ionizing radiation for all models. The abscissa is the
product of the effective ionizing photon rate $\Qphot = \fion\Qi$ and the rms
number density of the ionized gas
$\nrms = (\int_{\rm \mathcal{V}} \nion^2 dV/\int_{\rm \mathcal{V}} dV)^{1/2}$,
which are often accessible to observers via free-free radio continuum and/or
nebular emission lines. We take the integration volume $\mathcal{V}$ as a sphere
around the cluster center that encloses 99\% of $\int \nion^2 dV$ over the whole
domain.\footnote{The value of $\nrms$ can vary by a factor of $\sim$2 if we
  choose the integration volume that encloses 90\% or 99.9\% of the value over
  the whole domain.} In each model, $\Qphot\nrms$ increases with time in the
early phase of evolution, but decreases as gas is removed by feedback in the
late phase. Thus, individual model tracks start at the left, evolve to the
right, and then return toward the left. Meanwhile, $\fesci$ tends to secularly
increase and $\fion$ to decrease with time, while $\fdusti$ starts small,
reaches a maximum, and then decreases again.

\citet{pet72} derived an analytic expression for the dust absorption fraction
for a uniform-density, embedded, spherical \HII\ region with a constant
dust-to-gas ratio \citep[see also][]{ino02}. Their predictions for $\fion$ and
$\fdusti$ ($= 1 - \fion$ since $\fesci\equiv 0$), both as functions of
$\Qphot\nrms$, are plotted as orange solid lines in Figure~\ref{f:Qeffn}(b) and
(c). This model predicts that the photon absorption is dominated by dust when
$\Qphot\nrms$ is very large. Also considering a spherical \HII\ region but
including radiation pressure on dust and solving for the dynamical equilibrium
radial profiles, \citet{dra11b} found that strong radiation pressure acting on
dusty gas creates a central cavity and an outer high-density shell. The
resulting absorption fractions are plotted as dashed lines in
Figure~\ref{f:Qeffn}. Because of the enhanced density in the outer
radiation-compressed shell in the \citet{dra11b} model, recombination raises the
neutral fraction and hydrogen can absorb a larger fraction of ionizing photons,
raising $\fion$ and lowering $\fdusti$ relative to the uniform model of
\citet{pet72}. In the limit of $\Qi\nrms\rightarrow \infty$, the \citet{dra11b}
model predicts $\fion\rightarrow 0.31$ and $\fdusti\rightarrow 0.69$ for the
parameters we adopt ($\beta = 1.41$, $\gamma=7.58$; see Eqs. (6) and (7) in
\citealt{kim16}). Although the \citet{dra11b} solutions were calculated under
the assumption of static equilibrium, we previously showed \citep{kim16} that
the interior structure of spherical \HII\ regions that are undergoing
pressure-driven expansion (with both radiation and gas pressure) are in good
agreement with the profiles predicted by \citet{dra11b}. For fixed $\Qi$,
$\nrms$ decreases over time; following the \citet{dra11b} solution for a
spherical, embedded \HII\ region, this would correspond to a decrease in
$\fdusti$ and increase in $\fion$ over time.

Our numerical results show that both $\fion$ and $\fdusti$ depend on the
evolutionary state and generally do not follow the trends
expected for spherical \HII\ regions. This is of course because (1) \HII\
regions in our simulations have highly non-uniform, non-spherical distributions
of gas and dust and (2) a significant fraction of photons can escape without
being caught by the dusty gas. Even in the embedded phase with $\fesci \ll 1$,
\HII\ regions in high-$\Sigmacl$ clouds have $\fion$ higher than the theoretical
predictions for given $\Qphot\nrms$. This is likely caused by turbulent mixing
that transports neutral gas to the interiors of \HII\ regions, making them
non-steady and out of ionization-recombination equilibrium.
We also note that in a system containing multiple sources with similar
individual
values of $\nrms$, $\fion$, and $\fdusti$,  the numerical curves
would appear to the right of the
analytic curves because the total $\Qphot$ would be a multiple of the individual
values.  However, this cannot account for the orders of magnitude shift to
the right relative to the single-source analytic $\fion$ curve.  
 Moreover, whereas $\fion$ would increase in
time for an expanding spherical \HII\ region, in fact $\fion$ decreases in
time for the simulations (because of escaping radiation).

Although the spherical analytic predictions for $\fion$ appear uncorrelated
with results from simulations, there is some resemblance between the 
analytic prediction and the numerical results for $\fdusti$, in that
the former marks the upper envelope of the latter's distribution. One
possible reason that this may not be entirely a coincidence is that
$\fdusti$ is greatest at a late stage when the \HII\ region
most resembles an idealized shell-bounded Str\"{o}mgren sphere with a central
source.

The hydrogen absorption fraction $\fion$ of Galactic \HII\ regions has been
estimated by \citet{ino01}, \citet{ino02}, and \cite{bin18}. \citet{ino01}
estimated $\fion$ of Galactic \HII\ regions using the model of \citet{pet72} for
dusty \HII\ regions. For Galctic ultracompact and compact \HII\ regions,
\citet{ino02} derived a relation between $\fion$ and the ratio between the total
IR and unobscured H$\alpha$ (or free-free) fluxes assuming that UV photons
absorbed by dust grains are re-emitted in IR. \citet{bin18} estimated $\fion$ of
massive star forming regions by taking the ratio between the ionizing photon
rate obtained from the Planck free-free emission and the total ionizing photon
rate estimated from known massive stellar content. Their estimated values of
$\fion$ are plotted as open diamonds \citep{ino01}, squares \citep{ino02}, and
circles \cite{bin18} in Figure~\ref{f:Qeffn}(b). We note that \citet{ino01} and
\citet{ino02} did not account for radiation escape, so that the observed $\fion$
corresponds to an upper limit on the real hydrogen absorption fraction.
\citet{dra11b} attributed the range of observed $\fion$ to the variations in the
dust-to-gas ratio. Since $\fion$ varies during evolution of a star-forming GMC
in our simulations, the observed diversity of $\fion$ may also reflect that the
observed \HII\ regions are at a different evolutionary stage.

\subsection{Similarity between $\fescn$ and $\fesci$}\label{s:similarity}

\begin{figure*}[t!]
  \epsscale{0.95}\plotone{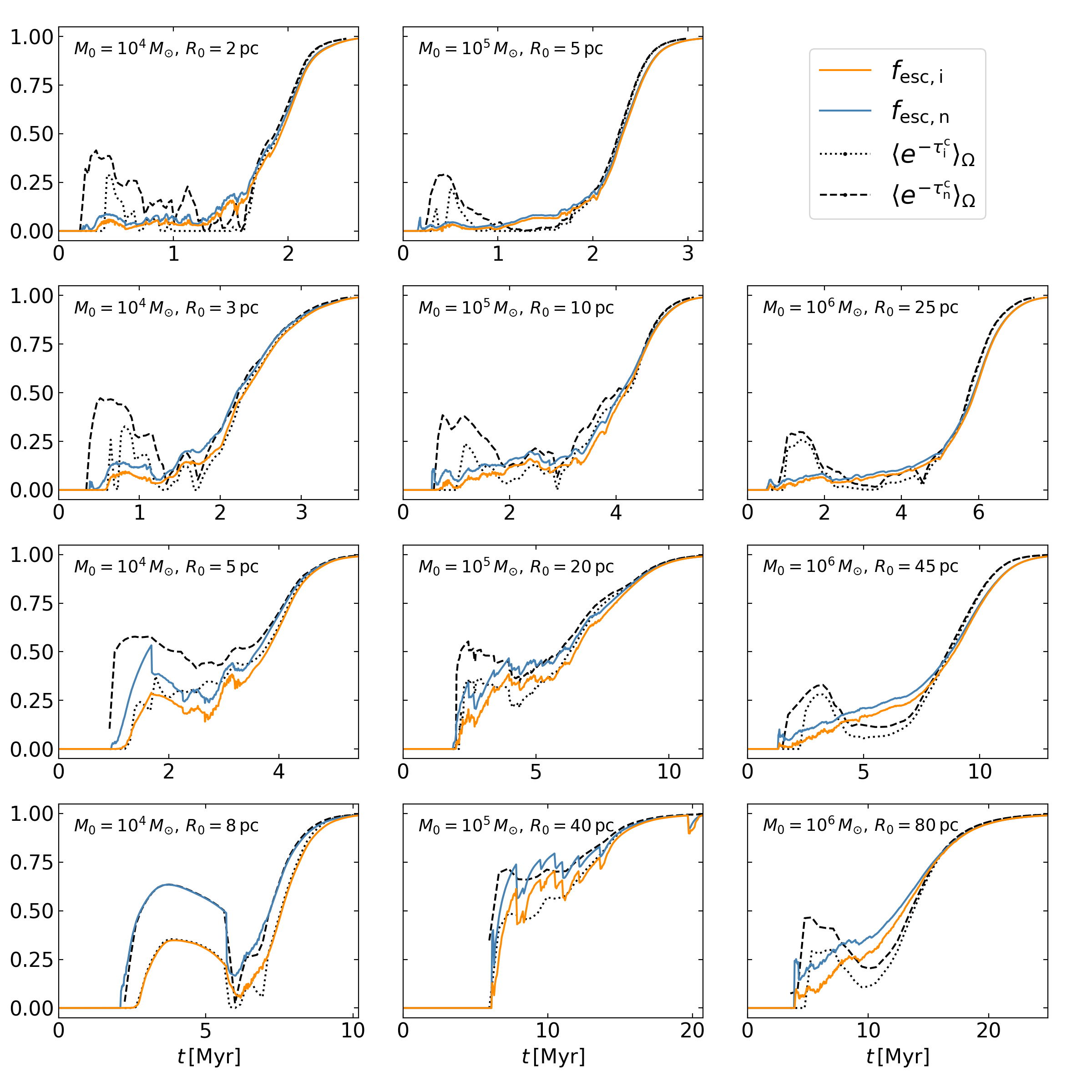}
  \caption{Evolution of the instantaneous escape fraction of ionizing ($\fesci$,
    orange) and non-ionizing ($\fescn$, blue) radiation for selected
    models whose mass and radius are specified in each panel. The dotted and
    dashed lines in black draw the escape fraction of ionizing radiation
    $\langle \exp(-\tauc_{\rm i}) \rangle_{\Omega}$ and non-ionizing radiation
    $\langle \exp(-\tauc_{\rm n}) \rangle_{\Omega}$, respectively, measured from
    the center of the stellar luminosity (see Section
    \ref{s:PDF-Omega}).}\label{f:fesc_comp}
\end{figure*}

Figure~\ref{f:fesc_comp} plots evolution of the escape fraction of ionizing
(orange) and non-ionizing (blue) radiation for selected models. Notably, the
difference between $\fesci$ and $\fescn$ is small or only modest. This is also
clear from the comparison of the angular distributions of $\exp(-\tauc_{\rm i})$
and $\exp(-\tauc_{\rm n})$, shown in the bottom row of Figures~\ref{f:snapshot1}
and \ref{f:snapshot2}. Although the covering fraction of optically thick
clumps/filaments to ionizing radiation is slightly enhanced relative to the
non-ionizing counterpart owing to the presence of the recombining gas in
photoevaporation flows, overall they appear quite similar. The reason that
$\fescn$ and $\fesci$ appear so similar is that both ionizing and non-ionizing
photons escape through low-density channels in which the gas is almost fully
ionized and the \HII\ region is density bounded.

For our models, the difference between the escape probabilities for a single
line of sight is
$e^{-\tau_{\rm n}} - e^{-\tau_{\rm i}} = e^{-\tau_{\rm n}}(1 - e^{-\tau_{\rm
    ph}})$, where $\tau_{\rm n} = \tau_{\rm d}= \int \nH\sigma_{\rm d} d\ell$ is
the dust optical depth (assumed to be the same for FUV and EUV) and
$\tau_{\rm ph} = \int\nHI\sigma_{\rm ph} d\ell
= \langle \xn\rangle \tfrac{\sigma_{\rm ph}}{\sigma_{\rm d}}\tau_{\rm n} $ is
the optical depth due to photoionization of neutral hydrogen, for the mean
neutral fraction $\langle \xn \rangle$. Note that the difference is bounded
above by $e^{-\tau_{\rm n}}$, and also bounded above by $\tau_{\rm ph}$. Both of
these upper limits can help to explain why $\fescn$ and $\fesci$ are similar, in
different circumstances.

If an \HII\ region is ionization bounded ($e^{-\tau_{\rm ph}} \ll 1$) along most
sightlines (high covering fraction of neutral gas), the difference between the
escape fractions of ionizing and non-ionizing radiation is determined by the
dust optical depth. In this case, the escape fractions of ionizing and
non-ionizing radiation are small and almost equal as long as
$\tau_{\rm d} \gg 1$ along most sightlines. This explains why $\fescn$ and
$\fesci$ are nearly identical in the highest-$\Sigmacl$ clouds at early times
(models {\tt M1E4R02}, {\tt M1E5R05}, and {\tt M1E6R25}). However, in
low-$\Sigmacl$ clouds at early times, a non-negligible fraction of non-ionizing
photons can escape through sightlines along which the \HII\ region is
ionization-bounded ($\tau_{\rm ph} \gg 1$) but has $\tau_{\rm d} \lesssim 1$.
This can explain noticeable differences between $\fesci$ and $\fescn$ at early
times in low-$\Sigmacl$ clouds (models {\tt M1E4R08}, {\tt M1E4R05}, {\tt
  M1E5R20}, {\tt M1E5R40}, and {\tt M1E6R80}).

At late stages of evolution for all models, the \HII\ region breaks out and
becomes density bounded along most sightlines
(high covering fraction of ionized gas), i.e. $\tau_{\rm ph} \lesssim 1$.  In
this circumstance, since $e^{-\tau_{\rm n}} - e^{-\tau_{\rm i}} < \tau_{\rm ph}$,
the difference between $\fescn$ and $\fesci$ will depend on the value of
$\tau_{\rm ph}$, which depends in turn on the ionization fraction.  

Quantitatively, for low density gas exposed to ionizing radiation,
$\langle \xn \rangle$ is close to the equilibrium value $x_{\rm n,eq}$
determined by photoionization-recombination balance $\mathcal{I}
\approx \mathcal{R}$, where $\mathcal{R} = \alphaB \nion \nelec \approx
\alphaB(1 - \xn)^2\nH^2$ is the local recombination rate, with
$\alphaB = 3.03\times 10^{-13}\cm^{3}\second^{-1}$ being the case B
recombination coefficient. Solving for $x_{\rm n,eq}\; (\ll 1)$ gives
\begin{equation}
 x_{\rm n,eq} \approx \dfrac{x_{\rm n,eq}}{(1 - x_{\rm n,eq})^2} 
 = \dfrac{\alphaB\nH}{c\sigma_{\rm ph} \mathcal{E}_{\rm i}/(h\nu_{\rm
     i})}\,. \label{eq:xneq1}
\end{equation}
Note that $x_{\rm n,eq}$ is inversely proportional to the local
ionization parameter $\mathcal{E}_{\rm i}/(\nH h\nu_{\rm i})$.
On directions that are density-bounded, the
radius $R_{\rm }$ is less than the Str\"omgren radius so that
$\nH < [(3Q_{\rm phot,i})/(4\pi\alphaB R_{\rm }^3)]^{1/2}$. 
Taking $\mathcal{E}_{\rm i} \sim h\nu_{\rm i}\Qi/(4\pi c R_{\rm }^2)$ along 
ionized directions, one can obtain
\begin{equation}
  x_{\rm n,eq} \lesssim \dfrac{1}{\sigma_{\rm ph}}\left(\dfrac{12\pi\alphaB R_{\rm }}{Q_{\rm phot,i}}\right)^{1/2}
\sim  3.0\times 10^{-4} Q_{\rm
    phot,i,49}^{-1/2} R_{\rm 2}^{1/2} \label{eq:xneq2}
\end{equation}
with $Q_{\rm phot,i,49} = Q_{\rm phot,i}/(10^{49}\second^{-1})$ and
$R_{\rm 2} = R_{\rm }/(10^2\pc)$.
We then have
\begin{equation}
  \tau_{\rm ph} \approx 5.4\times 10^3 \langle x_{\rm n,eq} \rangle \tau_{\rm d}
  \lesssim 1.6 Q_{\rm  phot,i,49}^{-1/2} R_{\rm 2}^{1/2}\tau_{\rm d}.
  \label{eq:taupheq}
\end{equation}  
 Equations~\eqref{eq:xneq2} and \eqref{eq:taupheq} suggest that
 optically observed \HII\ regions ($\tau_{\rm d} \lesssim 1$) which are
 bright and compact 
($Q_{\rm phot,i,49}^{1/2}R_{\rm 2}^{-1/2} \gg 1$) would have very low 
 neutral fraction $\xn \ll 1$ along density-bounded directions 
 and also $\tau_{\rm ph} \ll 1$, and thus would 
have $\fesci \approx \fescn = \langle e^{-\tau_{\rm d}}\rangle$. 
This explains our result that $\fesci \approx \fescn$ at late times in
all models.

\subsection{Cumulative Escape Fraction Before First Supernovae}\label{s:fesccum}

\begin{figure}[!t]
  \epsscale{1.15}\plotone{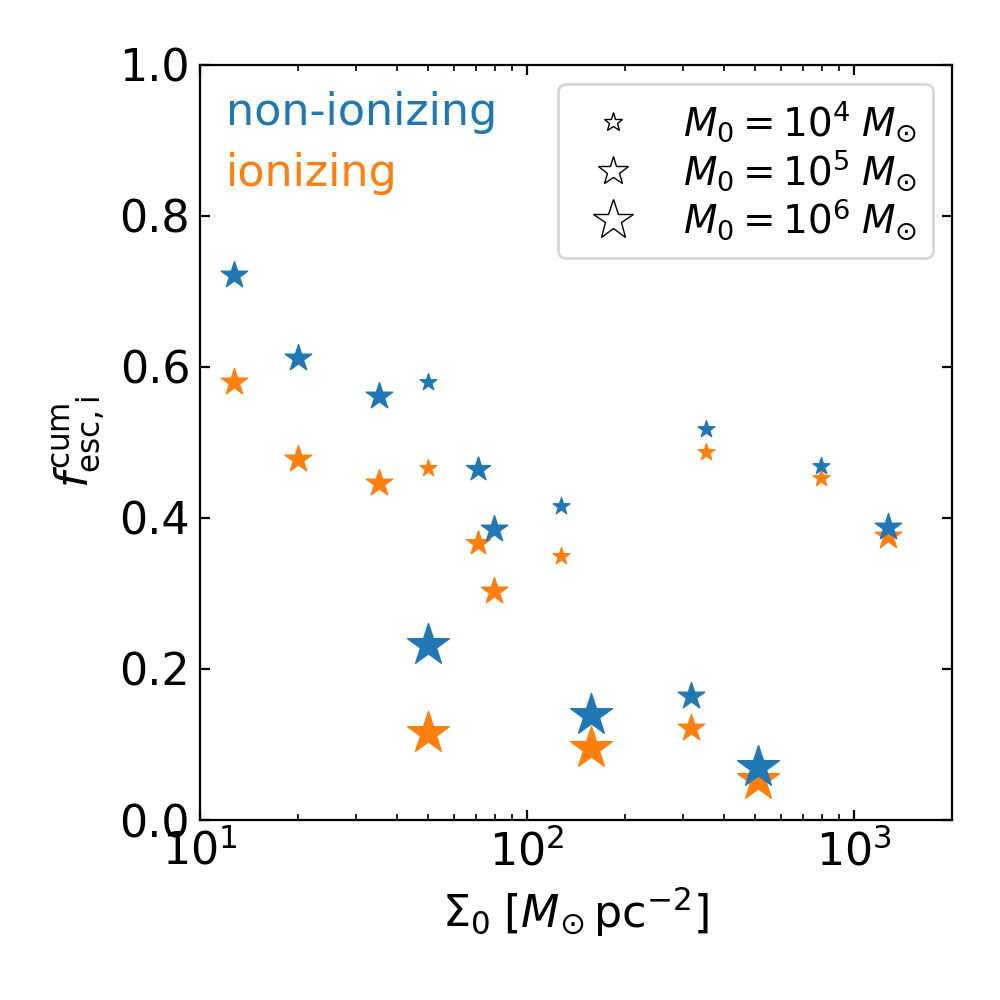}
  \caption{Cumulative escape fractions of ionizing (orange) and non-ionizing
    (blue) radiation up to time $t'=3 \Myr$ after the first star formation,
    plotted against the initial cloud surface density $\Sigmacl$.
    Although the cumulative escape fraction tends to
    decrease with increasing $\Sigmacl$,
    dense clouds with $t_{\rm dest} < 3\Myr$ have relatively high cumulative
    escape fractions because of rapid destruction.}\label{f:fcum_Sigma}
\end{figure}

Although our simulations do not account for the time variation of UV luminosity
due to stellar evolution, it is worth examining the cumulative fraction of UV
photons that escape from the natal cloud up to the time when the first supernova
explosions would occur. After this time, impact of SN blasts would affect the
cloud structure, and the ionizing photon production rate would drop
considerably. Figure~\ref{f:fcum_Sigma} plots the cumulative escape fraction of
ionizing (orange) and non-ionizing (blue) photons at $t'=t-t_{*,0}= 3 \Myr$ as a
function of $\Sigmacl$. These values together with the cumulative dust
absorption fractions are also listed in Columns 9--13 of Table~\ref{t:result}.

In general, both $f_{\rm esc,i}^{\rm cum}$ and $f_{\rm esc,i}^{\rm cum}$ tend to
decrease with increasing $\Sigmacl$, except for models {\tt M1E4R02}, {\tt
  M1E4R03}, and {\tt M1E5R05} which have
$f_{\rm esc,i}^{\rm cum}, f_{\rm esc,n}^{\rm cum} \sim 0.4$--$0.5$. These dense
clouds have a short evolutionary timescale, with $t_{\rm dest} < 3\Myr$ (see
Column 7 of Table~\ref{t:result}). In contrast, massive clouds
($\Mcl = 10^6 \Msun$) have a relatively long evolutionary time, and only a tiny
fraction of the initial gas mass has been ejected by radiation feedback at
$t'=3\Myr$ (see Figure 15 in \citetalias{kim18}), leading to very low cumulative
escape fractions. Supernova feedback is expected to play a greater role than
radiation feedback in destroying these massive clouds. Destruction of
  these massive clouds by SNe at early times would also increase
  $\fesc^{\rm cum}$ above what is shown in Figure~\ref{f:fcum_Sigma} and listed
  in Table~\ref{t:result}.

\section{Escape Fraction vs.\ Optical Depth Distribution}\label{s:PDF}

The escape fraction is intrinsically linked to the distribution of optical depth
around the sources that emit radiation. In this section, we will first calculate
the solid-angle probability distribution function (PDF) of the optical depth as
seen from the luminosity center of the sources, and show that its mean and
dispersion can be used to predict the escape fraction. Next, we calculate the
area PDF of the optical depth projected through the whole cloud as seen by an
external observer, and explore ways to estimate the escape fraction using this
area PDF. We focus mainly on the escape fraction of non-ionizing radiation since
this is determined by the dust optical depth distribution, which can be traced
observationally using far-IR thermal dust emission or near-IR extinction mapping
\citep[e.g.,][]{lom14}. As demonstrated in Section \ref{s:similarity},
$\fesci$ is expected to be similar to $\fescn$.

\subsection{Solid Angle-Weighted PDF of Optical Depth}\label{s:PDF-Omega}

We first provide a general framework to consider escape of radiation
from an inhomogeneous cloud, and then we turn to results from our
simulations.

For an isotropically emitting point source, the escape fraction of radiation is
determined by the solid angle distribution of the optical depth measured from
the source. Consider a point source embedded in an isolated dusty cloud with
mass $M$ and constant dust opacity per unit mass
$\kappad = \sigma_{\rm d}/\mu_{\rm H}$. The dust optical depth averaged over the
solid angle $\Omega$ is
\begin{equation}\label{eq:taucmean}
  \taucmean = \dfrac{\iint \rho(r,\Omega)\kappad dr d\Omega}{\int d\Omega}
  \equiv \kappad \langle \Sigma^{\rm c}\rangle_{\Omega},
\end{equation}
where $\rho(r, \Omega)$ is the gas density and
$\langle \Sigma^{\rm c}\rangle_{\Omega} = M/(4\pi\bar{r}^{2})$ is the
characteristic surface density of the circumsource material with
$\bar{r} \equiv (\int r^{-2} dM/\int dM)^{-1/2}$. Here, the superscripts ``c''
again indicate measurements of circumsource material relative to the cluster
center. Let $P_{\Omega}(\ln\,\tauc)d\ln\,\tauc = d\Omega/(4\pi)$ denote the
fraction of the whole solid angle covered by sightlines with the logarithm of
the dust optical depth in the range between $\ln\,\tauc$ and
$\ln\,\tauc + d\ln\,\tauc$. The escape fraction of non-ionizing radiation is
then given by
\begin{equation}\label{eq:solid_pdf}
  \langle e^{-\tauc} \rangle_{\Omega} = \int e^{-\tauc}
  P_{\Omega}(\ln\,\tauc)\,d\ln\,\tauc \geq e^{-\langle \tauc \rangle_{\Omega}}\,,
\end{equation}
where the inequality follows from
$\langle e^{-\tauc} \rangle_{\Omega} = e^{-\langle \tauc
  \rangle_{\Omega}}\langle e^{-\tauc + \langle \tauc \rangle_{\Omega}}\rangle$ and
$\langle e^{-\tauc + \langle \tauc\rangle_{\Omega}} \rangle \geq \langle (1 -
\tauc + \langle \tauc \rangle_{\Omega}\rangle_{\Omega} = 1$. Note that this
inequality holds independent of the functional form for the PDF of $\tauc$.
Equation \eqref{eq:solid_pdf} states that the true escape fraction
$\langle e^{-\tauc} \rangle_{\Omega}$ is always greater than or equal to the
naive estimate $e^{-\langle \tauc \rangle_{\Omega}}$ based on the mean optical
depth.

\begin{figure}[!t]
  \epsscale{1.15}\plotone{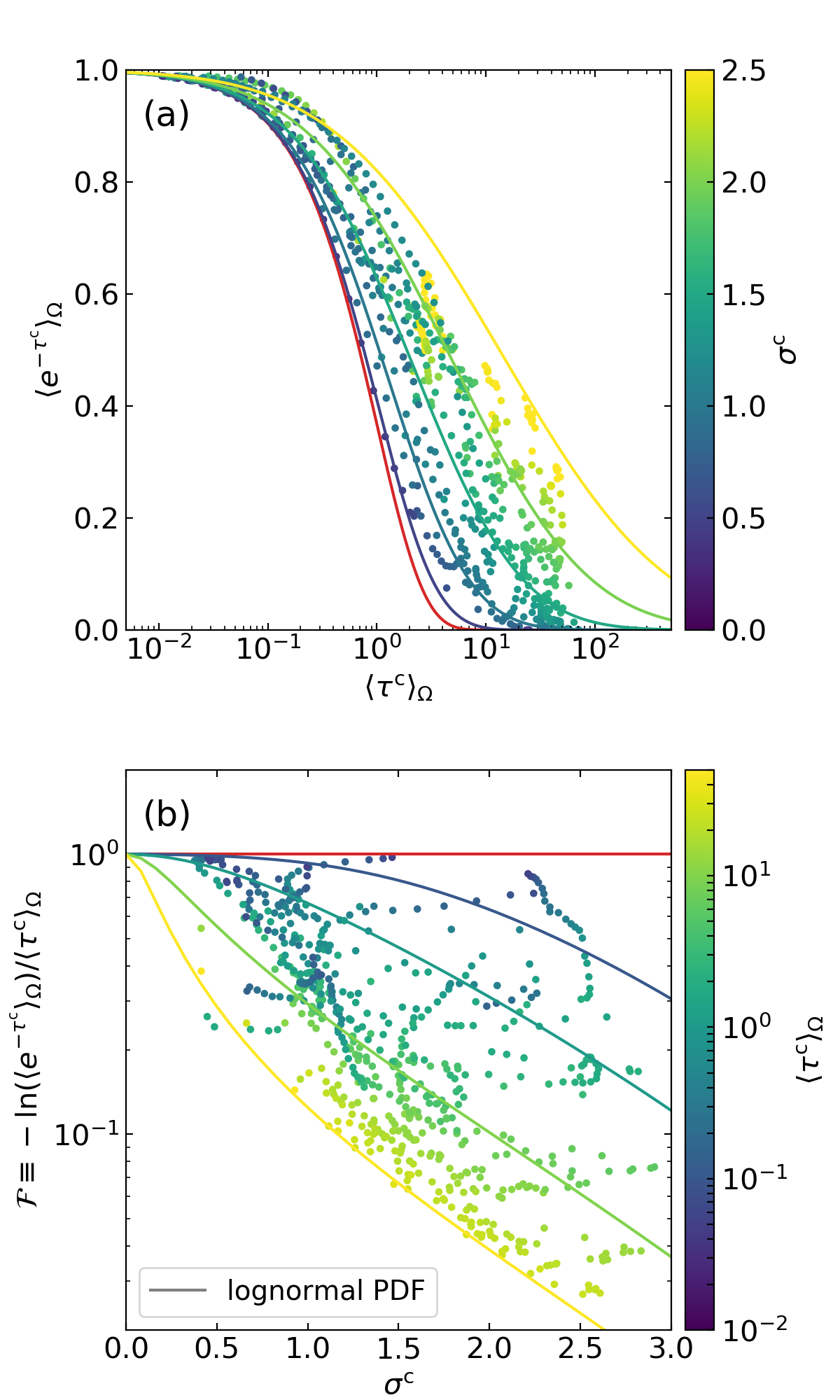}
  \caption{(a) Escape fraction of non-ionizing radiation as a function of the
    solid-angle averaged optical depth $\langle \tauc \rangle_\Omega$ seen from
    the source. With solid lines we show expectations based on lognormal
    distributions of the optical depth, with standard deviation
    $\sigmac = 0, 0.5, \cdots, 2.5$ from left to right. (b) The reduction factor
    $\mathcal{F}$ defined as the ratio of the effective optical depth
    $\taueff=-\ln \langle e^{-\tauc} \rangle_\Omega$ to $\taucmean$ as a
    function of $\sigmac$. The solid lines indicate the reduction factor
    expected for lognormal distributions with $\taucmean = 0, 0.1, 1, 10, 50$,
    from top to bottom. Small circles in (a) show characteristic escape
    fraction 
    and in (b) show characteristic reduction factor as measured from each
    simulation snapshot, where for this purpose we assume that all the sources
    are gathered at the center of luminosity. Colors correspond to measured
    value of $\sigmac$ in (a) and $\langle \tau^c \rangle_\Omega$ in
    (b).}\label{f:lognorm}
\end{figure}

A broad distribution of the optical depth can make
$\langle e^{-\tauc} \rangle_{\Omega}$ much larger than the naive estimate. To
demonstrate this, we consider an idealized situation in which $P_{\Omega}$
follows a lognormal distribution
\begin{equation}\label{e:PLN}
  P_{\Omega, {\rm LN}}(\ln\tauc; \muc, \sigmac) \equiv 
  \dfrac{1}{\sqrt{2\pi}\sigmac}\exp\left[-\frac{(\ln\tauc - \mu^{\rm
      c})^2}{2(\sigmac)^2}\right],
\end{equation}
with the mean $\muc = \langle \ln\tauc \rangle_{\Omega}$ and the standard
deviation $\sigmac = \langle (\ln\tauc - \muc)^2 \rangle_{\Omega}^{1/2}$. The
mean optical depth is then given by $\taucmean = e^{\muc + (\sigmac)^2/2}$.

In Figure~\ref{f:lognorm}(a) we plot as solid lines the escape fraction
$\langle e^{-\tauc} \rangle_{\Omega}$ as a function of $\taucmean$. All curves
are based on lognormal distributions, and each line is colored by its value of
$\sigmac$, given by $\sigmac = 0, 0.5, \cdots, 2.5$ from left to right. For
$\sigmac \rightarrow 0$, $P_{\Omega}(\ln \tauc)$ becomes a delta function and
$\langle e^{-\tauc} \rangle_{\Omega} \rightarrow e^{-\langle \tauc
  \rangle_{\Omega}}$, plotted as the red solid line. Note that the escape
fraction is close to unity regardless of $\sigmac$ when $\taucmean \ll 1$. For
$\taucmean\gtrsim1$, however, nonzero $\sigmac$ can boost the escape fraction by
a large factor relative to the $\sigmac=0$ case. For example, when
$\taucmean = 10$, the escape fraction is 0.18 when $\sigmac=1.5$, which is
$4000$ times higher than the value $e^{-10} \approx 4.5 \times 10^{-5}$ that
applies when $\sigmac=0$, since a significant fraction of the sky has
$\tauc \lesssim 1$ when the cloud is nonuniform.

The boost of the escape fraction due to inhomogeneous gas distributions around
sources corresponds to a reduction in the effective optical depth,
$-\ln (\langle e^{-\tauc} \rangle_{\Omega})$. We define the reduction factor
\begin{equation}\label{eq:Fred1}
 \mathcal{F} \equiv - \frac{\ln (\langle e^{-\tauc} \rangle_{\Omega})}{\taucmean} \le 1,
\end{equation}
which quantifies how much the effective optical depth is reduced relative to the
mean optical depth. In Figure~\ref{f:lognorm}(b) we plot with solid lines
$\mathcal{F}$ as a
function of $\sigmac$ for a lognormal PDF with several different values of
$\taucmean$. Curves are colored to indicate the value of
$\taucmean = 0, 0.1, 1, 10, 50$, from top to bottom. Again, the reduction factor
is close to unity regardless of $\sigmac$ for $\taucmean \ll 1$, but
$\mathcal{F}$ can be as small as 0.1 when $\taucmean \sim 10$ and
$\sigmac \sim 2$.

We now turn our attention to $P_{\Omega}(\ln \tau_{\rm n}^{\rm c})$ for our
simulation data. For the purpose of measuring a characteristic escape fraction
from the cloud in each simulation snapshot, we assume that all radiation is
emitted from a single point source located at the stellar center of luminosity
$\mathbf{r}_{\rm CL} = \sum_k L_k \mathbf{r}_k/\sum_k L_k$. We use a trilinear
interpolation to remap the density fields $\nH$ and $\nHI$ from the Cartesian
onto a spherical grid with $(N_r, N_{\theta}, N_{\phi}) = (128, 128, 256)$ zones
centered at $\mathbf{r}_{\rm CL}$. We set the radial grid spacing to
$\Delta r = (2\Rcl - |\mathbf{r}_{\rm CL}|)/N_r$, and calculate the optical
depth measured from $\mathbf{r}_{\rm CL}$.

In Figure~\ref{f:lognorm}, for all model snapshots we overlay as filled circles
(a) the escape fraction of non-ionizing radiation as seen from the center of
luminosity against the solid angle-averaged dust optical depth, and (b) the
optical-depth reduction factor as a function of the standard deviation of the
raw PDFs. As expected, all the data for the escape fraction measurements lie
above the red line in (a), corresponding to
$\langle e^{-\tau^{\rm c}}\rangle_\Omega = e^{-\taucmean}$, due to finite width
of the PDFs. The reduction factor becomes smaller with increasing $\taucmean$
and $\sigmac$, which is also qualitatively consistent with the lognormal PDF
prediction.

\begin{figure*}[!t]
  \epsscale{1.15}\plotone{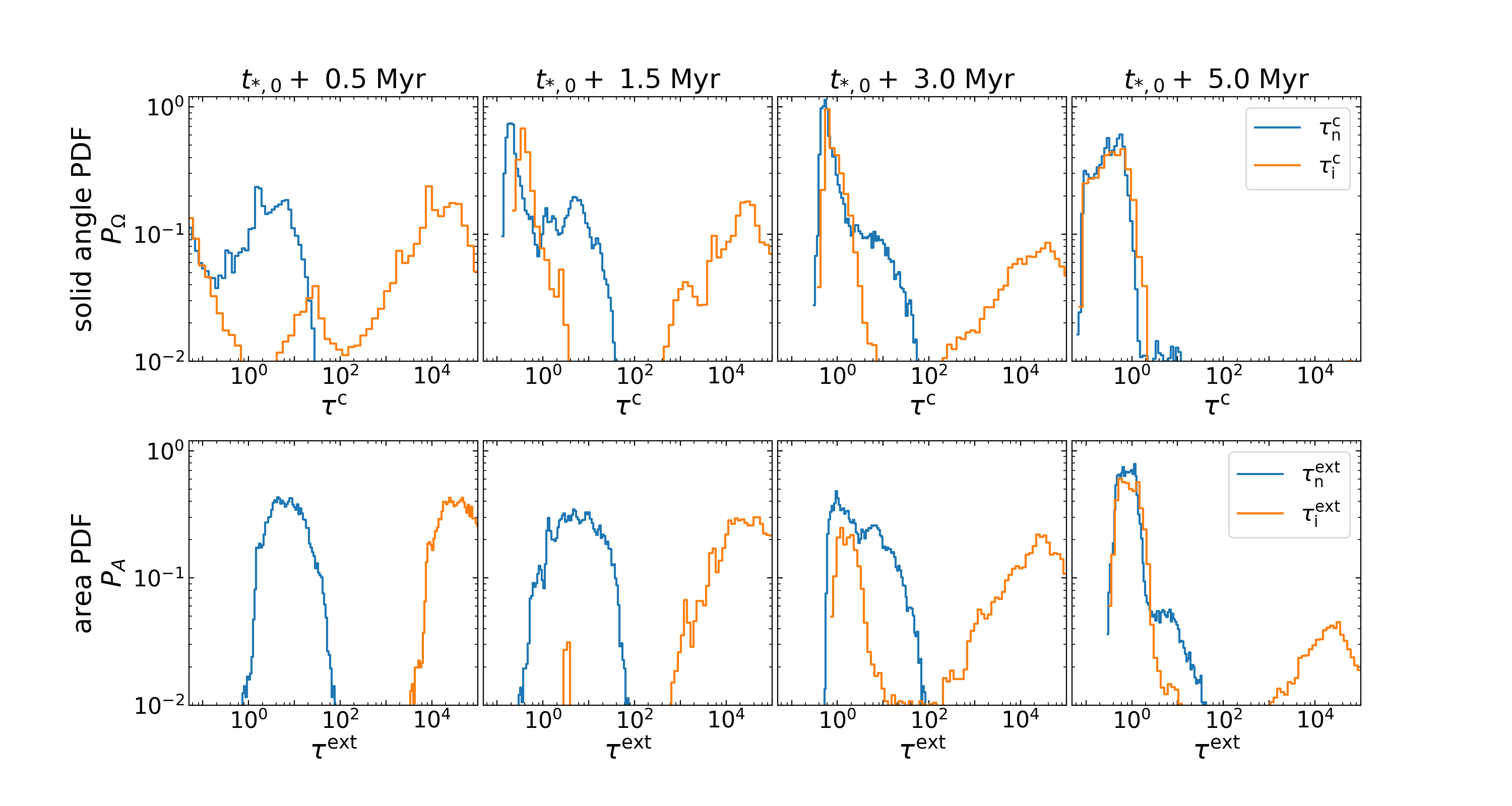}
  \caption{(Top) Solid-angle-weighted PDFs of the optical depth for non-ionizing
    ($\tau_{\rm d}^{\rm c}$, blue) and ionizing ($\tau_{\rm i}^{\rm c}$, orange)
    radiation measured from the stellar center of luminosity for the fiducial
    model at $0.5$, $1.5$, $3$, and $5 \Myr$ after the first epoch of star formation.
    (Bottom) Area-weighted PDFs of the optical depth projected along the three
    principal axes for non-ionizing ($\taud^{\rm ext}$, blue) and ionizing
    ($\taui^{\rm ext}$, orange) radiation within the half-mass radius
    $\Rh$.}\label{f:pdf}
\end{figure*}

The top row of Figure~\ref{f:pdf} plots the solid-angle PDFs of the
optical depth for non-ionizing (blue) and ionizing (orange) radiation,
as measured from the center of luminosity for the fiducial model at
the four different times shown in Figure~\ref{f:snapshot1}.  The PDFs,
in general, do not look like lognormal distributions, with multiple
peaks and shoulders associated with low-density holes and dense
neutral clumps. Except at very late times, the solid-angle PDF for
ionizing radiation is typically bimodal, while for non-ionizing
radiation the solid-angle PDFs are unimodal.

\begin{figure}[!t]
  \epsscale{1.15}\plotone{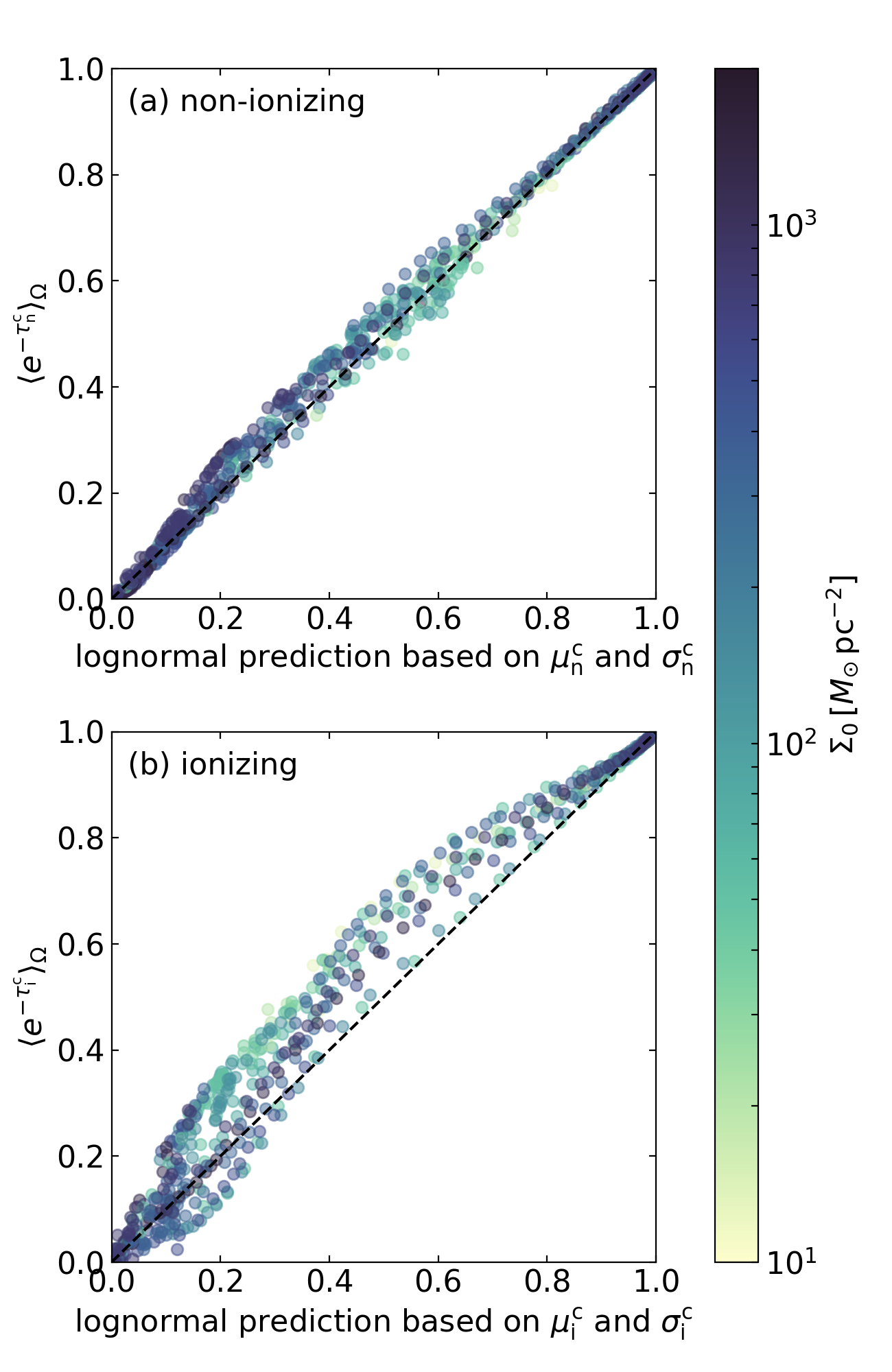}
  \caption{Comparison of the true escape fraction as seen from the cluster
    center with an estimated escape fraction for (a) non-ionizing and (b)
    ionizing radiation, for all simulation snapshots.  The true escape fraction
    $\langle e^{-\tau^{\rm c}}\rangle_\Omega$ on the ordinate is calculated from
    Equation~\eqref{eq:solid_pdf} using the raw PDF for $P_\Omega$.   The estimated
    escape fraction on the abscissa uses Equation (\ref{e:PLN}) for
    $P_\Omega$, with the mean (${\mu}^{\rm c}$)
    and width (${\sigma}^{\rm c}$) as measured from the raw solid-angle PDF.
    }\label{f:fesc-lognorm}
\end{figure}

For each simulation snapshot in all models, we measure the mean $\mu^{\rm c}$
and standard deviation $\sigma^{\rm c}$ from the raw PDFs, for both ionizing and
non-ionizing radiation. We then calculate what the escape fraction would be
using Equation \eqref{eq:solid_pdf} with a lognormal (Equation \ref{e:PLN}) for
$P_\Omega(\ln \tauc)$, using the measured $\mu^{\rm c}$ and $\sigma^{\rm c}$
values. We also directly evaluate Equation \eqref{eq:solid_pdf} using the raw
PDF for $P_\Omega(\ln \tauc)$ to obtain the true escape fraction
$\langle e^{-\tau^{\rm c}}\rangle_\Omega$ from the luminosity center.
Figure~\ref{f:fesc-lognorm} compares the true escape fractions with the
estimated escape fractions based on lognormals with the same $\mu^{\rm c}$ and
$\sigma^{\rm c}$, for all simulation snapshots. We show results for both (a)
non-ionizing and (b) ionizing radiation. The lognormal estimate agrees with the
raw escape fraction within $7\%$ for non-ionizing and within $20\%$ for ionizing
radiation. This suggests that quite a good estimate of the escape fraction can
be obtained given knowledge of the mean and variance in $\ln \tauc$. The
superiority of the estimated escape fraction for non-ionizing radiation compared
to ionizing radiation is not surprising, given that the former is typically
closer to a lognormal (as the example in Figure~\ref{f:pdf} shows), but our
results demonstrate that the escape fraction is insensitive to the detailed
functional form of the PDF.

The temporal evolution of $\langle e^{-\tau_{\rm i}^{\rm c}}\rangle_\Omega$ and
$\langle e^{-\tau_{\rm n}^{\rm c}}\rangle_\Omega$ for some selected models are
plotted as dotted and dashed lines in Figure~\ref{f:fesc_comp}. Overall, these
agree quite well with the luminosity-weighted escape fractions $\fescn$ and
$\fesci$, suggesting that distributed sources can be regarded as if they were
gathered at the luminosity center for the purpose of calculating the photon
escape fractions. We note that the predicted escape fraction from a single
source is somewhat larger than the actual escape fraction in the early phase of
evolution. This is because at early time sources are clustered in a few
widely-separated regions (e.g., leftmost column of Figure~\ref{f:snapshot1}) and
the luminosity center is located in a low-density void created by turbulence, in
which case the gas distribution around the luminosity center does not properly
represent the actual gas distributions surrounding individual sources.

\subsection{Area-Weighted PDF of Optical Depth}\label{s:PDF-area}

While the solid-angle PDF of the optical depth, $P_{\Omega}(\tauc)$, determines
the escape fraction, it is not directly available to an external observer. At
best, several individual line-of-sight values of $\tauc$ could be obtained from
spectral observations of stars within a cloud. If the \HII\ region is well
resolved, sampling of multiwavelength nebular spectra in sufficiently many
locations could also be used to estimate the distribution of optical depths
e.g., using the Balmer decrement method. Alternatively, given sufficient
resolving power, an external observer could use IR dust extinction or emission
maps to measure the area distribution, $P_A(\tau_{\rm n}^{\rm ext})$, of the
dust optical depth $\tau_{\rm n}^{\rm ext}$ projected on the plane of the sky.
Can the observer use $P_A(\tau_{\rm n}^{\rm ext})$ to obtain an estimate of the
escape fraction close to the real value? We explore this possibility below.

Since the area PDF defined over the entire domain depends on the box size, we
consider the gas only within the half-mass radius of the cloud as follows. We
first calculate the column-density weighted mean position of a gas cloud with
total mass $M_{\rm gas}$, and take it as the cloud center in the projected plane
of the sky (cross symbols in the top row of Figures~\ref{f:snapshot1} and
\ref{f:snapshot2}). Next, we draw a circle with the half-mass radius $\Rh$ about
the center that encloses 50\% of the total gas mass (dotted circles in the top
row of Figures~\ref{f:snapshot1} and \ref{f:snapshot2}). This allows us to
define the area-averaged surface density
$\langle \Sigma^{\rm ext}\rangle_A \equiv \tfrac{1}{2}{M_{\rm gas}}/(\pi\Rh^2)$
and the area-averaged non-ionizing (dust) optical depth
$\langle \tau_{\rm n}^{\rm ext} \rangle_{A} = \kappad\langle \Sigma^{\rm
  ext}\rangle_A$ within the half-mass radius.\footnote{Since we adopt a
    constant dust opacity per unit mass, the dust optical depth PDF is
    equivalent to the gas column density PDF. With
    $\kappa_{\rm d} = 500 \cm^{2}\gram^{-1} = 0.105 \pc^{2}\,\Msun^{-1}$, the
    unit dust optical depth corresponds to the gas column density of
    $9.59\Sunit$ or column density of hydrogen nuclei
    $N_H= 8.54 \times 10^{20} \cm^{-2}$.} We post-process all snapshots in the
time range ($t_{*,0},t_{\rm ej,99\%}$) at $0.05\tffcl$ interval, where
$t_{\rm ej,99\%}$ denotes the time at which 99\% of initial gas mass has been
ejected from the simulation domain.

The bottom row of Figure~\ref{f:pdf} plots as blue lines the area-weighted PDFs
of the dust optical depth within the half-mass radius in the fiducial model, at
four different times shown in Figure~\ref{f:snapshot1}. The PDFs along the three
principal ($x$, $y$, and $z$) axes are combined. At $t' = 0.5 \Myr$, the area
PDF is approximately lognormal since the density distribution is dominated by
supersonic turbulence \citep[e.g.,][]{mck07}. The fraction of area with
low $\tau_{\rm n}^{\rm ext}$ grows over time due to photoevaporation. At
$t'= 5 \Myr$, the area PDF exhibits a narrow width and a pronounced peak at
$\tau_{\rm n}^{\rm ext} \sim 1$.

In the bottom row of Figure~\ref{f:pdf}, we also plot as orange lines the
area-weighted PDFs of the projected optical depth for ionizing radiation
$\taui^{\rm ext} = \int (\nH\sigma_{\rm d} + \nHI\sigma_{\rm ph})\,dz$. At early
times, the area PDF of $\taui^{\rm ext}$ is largely similar in shape to the PDF
of $\tau_{\rm n}^{\rm ext}$ with a shift to the right by a factor
$\sigma_{\rm ph}/\sigma_{\rm d}$ because only a tiny fraction of sightlines are
optically thin and $\taui \sim \taud (\sigma_{\rm ph}/\sigma_{\rm d})$ along
most sightlines. At intermediate times, the area-weighted PDFs of
$\taui^{\rm ext}$ have two peaks and shoulders associated with neutral clumps
and ionized interclump gas. Later, the PDFs for $\taui$ and $\tau_{\rm n}$
become similar as neutral gas covers only a tiny fraction of the total area
within $\Rh$ and most sightlines have
$\xn \lesssim \sigma_{\rm d}/\sigma_{\rm ph}$.

\begin{figure}[!t]
  \epsscale{1.2}\plotone{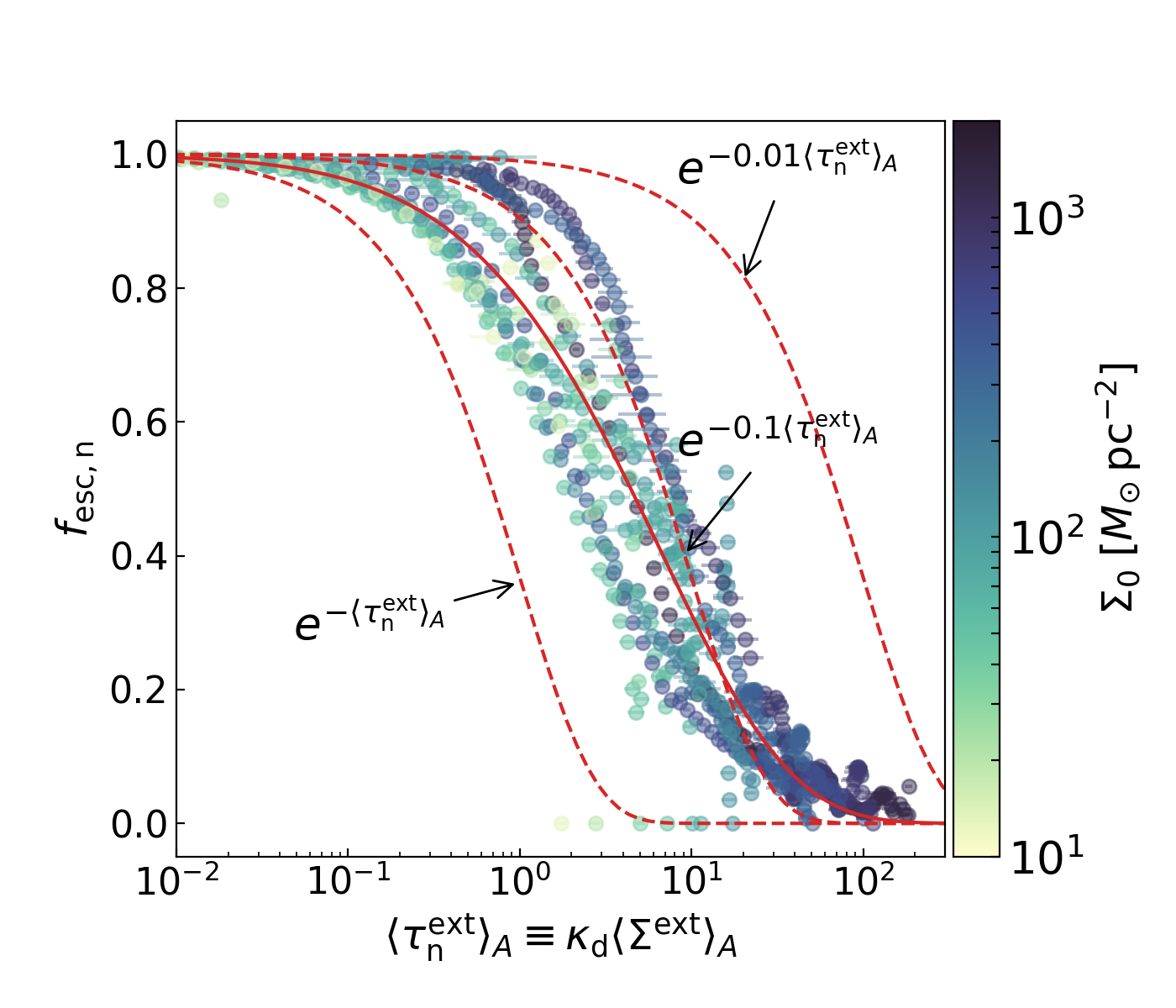}
  \caption{Instantaneous escape fraction $\fescn$ of non-ionizing radiation
    against the area-averaged dust optical depth
    $\langle \tau^{\rm ext}_{\rm n} \rangle_A$ within the half-mass radius for
    all models. The color of each dot indicates the initial surface density of
    the cloud. The horizontal bars represent the 1-sigma uncertainty based on
    $\langle \tau^{\rm ext}_{\rm n} \rangle_A$ measured along three different
    orientations. Red dashed lines provided for comparison correspond to
    $\fescn = e^{-\langle \tau_{\rm n}^{\rm ext}\rangle_A}$,
    $e^{-0.1\langle \tau_{\rm n}^{\rm ext}\rangle_A}$, and
    $e^{-0.01\langle \tau_{\rm n}^{\rm ext}\rangle_A}$ from left to right. This
    shows that the effective optical depth for escape of photons is much lower
    than the value implied by the mean column of gas in the cloud. The estimate
    of escape fraction $\fescn^{\rm est,1}$ is shown with a solid red curve (see
    Equation~\ref{eq:fesc_est1}).}\label{f:fescn-tauh}
\end{figure}

\subsubsection{Estimation of $\fescn$}\label{s:PDF-pred}

\begin{figure}[!t]
  \epsscale{1.2}\plotone{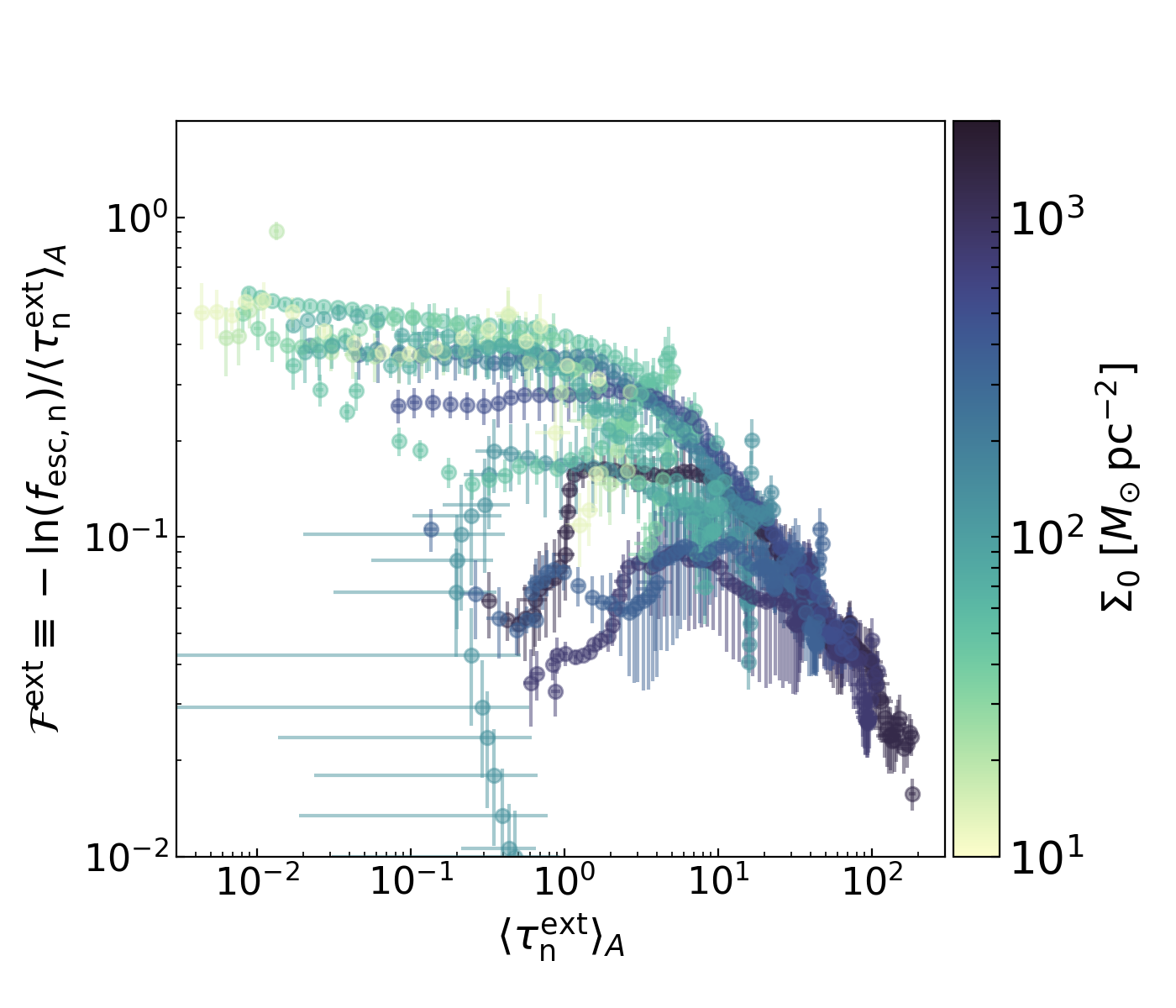}
  \caption{Area-averaged dust optical depth
    $\langle \tau_{\rm n}^{\rm ext}\rangle_A$ within the half-mass radius {\it
      vs.} the reduction factor
    $\mathcal{F}^{\rm ext} \equiv -\ln\,(\fescn)/\langle\tau_{\rm n}^{\rm
      ext}\rangle_A$ for an external observer for all models. The error bars
    show the standard deviation of measurements along three projection
    directions. The effective optical depth for photons escaping from
      embedded clusters is reduced by a factor $\sim 0.5$ (at low
      $\langle\tau_{\rm n}^{\rm ext}\rangle_A$) to $\sim 0.02$ (at high
      $\langle\tau_{\rm n}^{\rm ext}\rangle_A$) compared to the mean cloud
      optical depth $\langle\tau_{\rm n}^{\rm ext}\rangle_A$ seen by an external
      observer.}\label{f:Fred2}
\end{figure}

\begin{figure*}[!t]
  \epsscale{1.2}\plotone{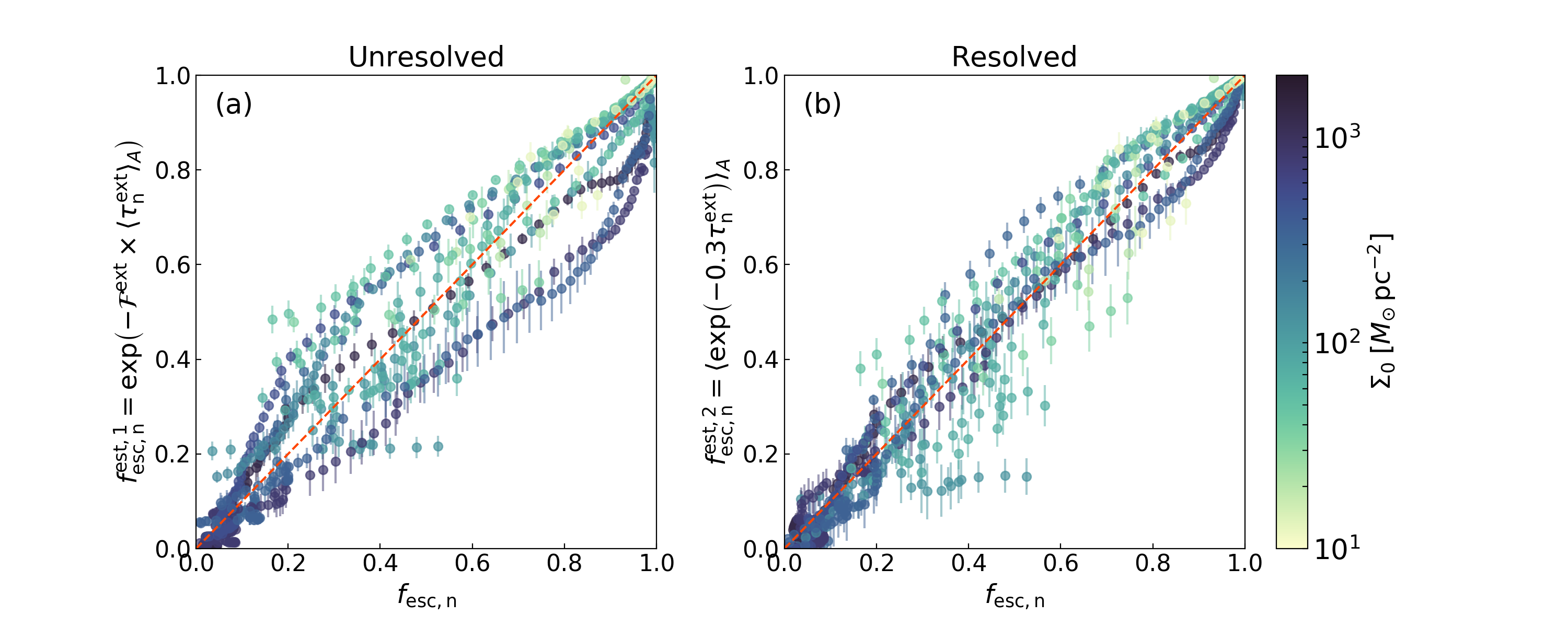}
  \caption{Actual escape fraction $\fescn$ of non-ionizing radiation {\it vs.}
    estimated escape fraction (a) based on the area-averaged optical depth
    $\fescn^{\rm est,1} = \exp(-\eta_1 \langle\tau_{\rm n}^{\rm ext}\rangle_A)$
    with
    $\eta_1 = 0.56/(1 + 1.25\langle \tau_{\rm n}^{\rm ext}\rangle_A^{0.48})$ and
    (b) the area-averaged escape fraction
    $f_{\rm esc,n}^{\rm est,2} =  \langle \exp(-\eta_2 \tau_{\rm n}^{\rm
      ext})\rangle_{A}$ with $\eta_2 = 0.30$. The dashed lines draw
    $\fescn / \fescn^{\rm est} = 1$.}\label{f:fesc_pred}
\end{figure*}

\paragraph{Marginally-resolved Cloud Case}
The simplest estimate of the escape fraction would make use of the mean column
of dust in a cloud, averaged over the aperture. This would be useful for a cloud
that is distant and not well resolved. Figure~\ref{f:fescn-tauh} plots the
instantaneous escape fractions $\fescn$ of non-ionizing radiation as a function
of the area-averaged optical depth $\langle \tau_{\rm n}^{\rm ext} \rangle_A$
for all models, with the color representing the initial cloud surface density.
The circles show the median value of $\langle \tau_{\rm n}^{\rm ext} \rangle_A$
measured along the three principal axes, while the horizontal bars indicate the
sample standard deviation, which is typically $\sim 20$--$30\%$ of
$\langle \tau_{\rm n}^{\rm ext} \rangle_A$. Note that all clouds start from
$\langle \Sigma^{\rm ext}\rangle_A \sim \Sigmacl$ and evolve towards a state
with $\langle \Sigma^{\rm ext}\rangle_A \rightarrow 0$ and
$\fescn \rightarrow 1$. For comparison, we plot as red dashed lines the simple
predictions assuming that the optical depth is equal to the mean value within
the half-mass radius, or is reduced by a factor of $10$ or $100$, i.e.
$e^{-\langle \tau_{\rm n}^{\rm ext}\rangle_A}$,
$e^{-0.1\langle \tau_{\rm n}^{\rm ext}\rangle_A}$, and
$e^{-0.01\langle \tau_{\rm n}^{\rm ext}\rangle_A}$ from left to right.

Figure~\ref{f:fescn-tauh} shows that the actual escape fraction is significantly
higher than the naive estimate $e^{-\langle \tau_{\rm n}^{\rm ext}\rangle_A}$.
The reason is twofold. First, using just a single mean optical depth does not
account for the variance associated with turbulence-driven structure, and leads
to an underestimate of the escape fraction for the reasons explained in
Section~\ref{s:PDF-Omega}. Second, even the area-averaged optical depth measured
by an external observer would be higher than the solid angle-averaged optical
depth measured by an internal observer located at the luminosity center. For
example, a uniform density sphere with radius $R$ and density $\rho$ has a
half-mass radius $\Rh=(1-2^{-2/3})^{1/2}R$ so that
$\langle\tau_{\rm n}^{\rm ext}\rangle_A = 1.8 \rho R \kappad$, nearly a factor
two larger than $\tauc_{\rm n}$. While not as extreme in turbulent clouds,
Figure~\ref{f:pdf} shows that the mean $\langle\tau_{\rm n}^{\rm ext}\rangle_A$
of the area PDF is larger than the mean
$\langle\tau_{\rm n}^{\rm c}\rangle_{\Omega}$ of the solid angle PDF.

The results in Figure~\ref{f:fescn-tauh} suggest that an approximate estimate of
the escape fraction may be obtained by applying an appropriate reduction factor
to $\langle\tau_{\rm n}^{\rm ext}\rangle_A$. Similar to
Equation~\eqref{eq:Fred1}, we define the reduction factor
$\mathcal{F}^{\rm ext} \equiv -\ln\,(\fescn)/\langle\tau_{\rm n}^{\rm
  ext}\rangle_A$ for an external observer. It tells us what the reduction in the
effective optical depth is relative to the area-averaged optical depth
$\langle\tau_{\rm n}^{\rm ext}\rangle_A$ and depends both on the geometric
distribution of gas and stars and (weakly) on the viewing angle of the observer.
Figure~\ref{f:Fred2} plots as circles $\mathcal{F}^{\rm ext}$ as a function of
$\langle\tau_{\rm n}^{\rm ext}\rangle_A$ with error bars indicating the standard
deviation of the values measured along three principal axes. The reduction tends
to be more significant for snapshots with larger
$\langle\tau_{\rm n}^{\rm ext}\rangle_A$, similar to the trend we found for
$\mathcal{F}$ in Figure~\ref{f:lognorm} (see also Figure~\ref{f:fred-alt} in
Appendix).\footnote{For dense and compact clouds (${\tt M1E4R02}$,
  ${\tt M1E4R03}$, ${\tt M1E4R05}$, ${\tt M1E5R05}$), $\mathcal{F}^{\rm ext}$ is
  small ($\lesssim 0.1$) at late evolutionary stage even when
  $\langle\tau_{\rm n}^{\rm ext}\rangle_A \sim 1$ and the density distribution
  is relatively smooth. This is because the gas cloud is offset significantly
  from the stellar center of luminosity; the escape fraction is close to unity
  due to small covering fraction.} In the limit
$\langle\tau_{\rm n}^{\rm ext}\rangle_A \ll 1$, the reduction factor tends to
the geometric correction factor $\sim 1/1.8$ for uniform density sphere.

Based on the above findings, we estimate the escape fraction as
\begin{equation}
  \fescn^{\rm est,1} = \exp\left(-\eta_1\langle\tau_{\rm n}^{\rm
        ext}\rangle_A\right)\,. \label{eq:fesc_est1}
\end{equation}
As an estimate of a correction factor $\mathcal{F}^{\rm ext}$ based only on
information that would be available for a marginally-resolved cloud, we adopt a
functional form
$\eta_1 = \tfrac{1/1.8}{1 + a\langle\tau_{\rm n}^{\rm ext}\rangle_A^b}$, with
constants $a$ and $b$ to be determined.\footnote{We find that this functional
  form approximates the reduction factor $\mathcal{F}$ for the lognormal
  solid-angle PDF with a given $\sigma^{\rm c}$ very well, giving results within
  a few percent for
  $10^{-2} < \langle \tau^{\rm c} \rangle_{\Omega} < 3\times 10^2$ and
  $0.5 < \sigmac < 3.0$.} The estimate in Equation~(\ref{eq:fesc_est1}) depends
only on $\langle\tau_{\rm n}^{\rm ext}\rangle_A$ and approaches
$\eta_1 \rightarrow 1/1.8$ for
$\langle\tau_{\rm n}^{\rm ext}\rangle_A \rightarrow 0$. We perform a least
square fit to find parameters $a=1.25$, $b=0.48$ that minimize the sum of
squared errors ($(\fescn - \fescn^{\rm est,1})^2$) compared to our simulation
results. Figure~\ref{f:fesc_pred}(a) compares the actual escape fraction
$\fescn$ of non-ionizing radiation and
$\fescn^{\rm est,1} = \exp\left(\tfrac{-0.56\langle\tau_{\rm n}^{\rm
      ext}\rangle_A}{1 + 1.25\langle\tau_{\rm n}^{\rm
      ext}\rangle_A^{0.48}}\right)$ for all models, with color corresponding to
the initial cloud surface density. This estimator predicts $\fescn$ within
$\sim 20\%$. The estimator of Equation~(\ref{eq:fesc_est1}) is also shown as a
red solid line in Figure~\ref{f:fescn-tauh}).

\paragraph{Resolved Cloud Case}
We have also tested a second method to estimate the escape fraction assuming
that the area PDF of $\tau_{\rm n}^{\rm ext}$ is available. In this approach,
one may estimate the escape fraction by taking the direct area average
\begin{equation}
  \fescn^{\rm est,2} = \langle \exp(-\eta_2 \tau_{\rm n}^{\rm ext})\rangle_A\,,
\label{eq:fest2}
\end{equation}
where $\eta_2$ is a constant correction factor. To find the optimal value of
$\eta_2$, we calculate the individual correction factor $\eta$ that gives
$\langle \exp(-\eta \tau_{\rm n}^{\rm ext}) \rangle_A = \fescn$ for each
snapshot of all simulations. The resulting $\eta$ values range between 0.1 and
0.5, with an average value of 0.30 and standard deviation of 0.14. We also adopt
the constant value of $\eta_2$ for all snapshots and find that $\eta_2 = 0.30$
minimizes the sum of the square of the differences
$(\fescn^{\rm est,2} - \fescn)^2$. Figure~\ref{f:fesc_pred}(b) compares
$\fescn^{\rm est,2}$ with the actual escape fraction, again showing that this
method predicts $\fescn$ within 20\%.

These results indicate that the two methods based on externally-observed mean
dust optical depth or optical depth distribution around a young star cluster can
be reliably used to infer the actual escape fraction from the cluster. Although
the largest errors are comparable for the two methods, the mean errors are
smaller using the second method. This implies that more accurate estimates of
the escape fraction may be obtained when the resolved dust distribution (or gas
distribution, with an assumed dust-to-gas value) can be measured.

\section{Summary and Discussion}\label{s:summary}

\subsection{Summary}

Stellar UV photons escaping from star-forming regions have a profound influence
on the ISM, especially its thermal and chemical state, together with
the resulting dynamical evolution including star formation.
Despite this importance, the escape fraction has not previously been
well characterized in observations, and theoretical predictions are also
lacking.
In this work, we have used a suite of radiation hydrodynamic simulations to
study the evolution of the escape fractions $\fescn$ and $\fesci$ for both
non-ionizing and ionizing radiation, and to analyze in detail how the escape
fraction depends on the dust optical depth distribution. Our simulations span a
range of physical conditions and include the effects of photoionization and
radiation pressure from UV radiation, but do not consider stellar evolution and
other forms of stellar feedback such as stellar winds or supernovae.

Utilizing the adaptive ray tracing module, we accurately follow the propagation
of ionizing and non-ionizing radiation from multiple sources and monitor
temporal evolution of the escape fraction, the dust absorption fraction, and the
hydrogen absorption fraction. We also explore how the escape fraction is related
to the solid-angle weighted distribution of the optical depth as seen from the
center of sources, and to the area-weighted distribution of the optical depth as
seen from outside the cloud. Based on our results, we propose two methods to
estimate the escape fraction from the observed optical depth in the plane of
sky.

Our key findings are summarized below.

\begin{enumerate}

\item {\textit{Temporal Evolution}}

  In all of our simulations, the escape fraction increases with time and becomes
  unity within a few free-fall times after the onset of feedback
  (Figures~\ref{f:fevol} and \ref{f:fesc_comp}). While clouds with low surface
  density are dispersed by radiation feedback rather quickly in a single
  free-fall time (Figures~\ref{f:snapshot1} and \ref{f:fevol}), \HII\ regions
  formed in clouds with high surface density spend a long embedded-phase
  ($\sim 2$--$3 \tffcl$) during which the escape fraction of ionizing radiation
  is small ($\fesci \lesssim 0.1$), while the hydrogen absorption fraction
  ($\fion$) remains high (Figures~\ref{f:snapshot2} and \ref{f:fevol}). Overall,
  $\fion$ decreases more-or-less monotonically with time, while the dust
  absorption fraction of ionizing photons ($\fdusti$) reaches a peak slightly
  before the cloud destruction and decreases in the late evolutionary phase
  (Figure~\ref{f:fevol}). The escape of both ionizing and non-ionizing radiation
  occurs mainly through low-density regions, along directions for which 
  the \HII\ region is
  density-bounded (Figures~\ref{f:snapshot1} and ~\ref{f:snapshot2}). As a
  result, the difference between the escape fraction of non-ionizing ($\fescn$)
  and ionizing radiation is quite small or only modest
  (Figure~\ref{f:fesc_comp}), with the dust absorption controlling the escape of
  UV radiation in the late phase.

\item {\textit{Comparison to Semi-analytic Models for Spherical, Embedded \HII\
      Regions}}

  Previous theoretical models for spherical, static, and embedded \HII\ regions
  with $\fesci=0$ have predicted that $\fion = \Qphot/\Qi$ increases (and
  $\fdusti$ decreases) with increasing $\Qphot\nrms$, where $\Qphot$ is the
  absorption rate of ionizing photons by hydrogen and $\nrms$ is the rms number
  density of ionized gas inside an \HII\ region. In our simulations, however,
  the relationship between $\fion$ (or $\fdusti$) and $\Qphot\nrms$ depends on
  the evolutionary phase of an \HII\ region, and deviates considerably from the
  theoretical predictions (Figure~\ref{f:Qeffn}). The discrepancy between the
  spherical model prediction and our numerical results is caused by the fact
  that \HII\ regions in our simulations are highly non-uniform and subject to
  loss of ionizing radiation through optically-thin holes, and that in
  time-dependent flows ionization rates can be enhanced by ``fresh'' neutral gas.
  The range of $\fion$ in our simulations is consistent with observed
  estimates in galactic \HII\ regions.  

\item {\textit{Cumulative Escape Fraction Before First Supernovae}}

  The cumulative escape fraction of ionizing photons ($f_{\rm esc,i}^{\rm cum}$)
  before the time of the first supernovae ($3\Myr$ after the onset of radiation
  feedback) ranges from 5\% to 58\% (Table~\ref{t:result}). The range of
  $f_{\rm esc,n}^{\rm cum}$ for non-ionizing photons is 7\% to 72\%. For fixed
  cloud mass, and both EUV and FUV, $f_{\rm esc}^{\rm cum}(3\Myr)$ tends to
  decrease with increasing $\Sigmacl$ (Figure~\ref{f:fcum_Sigma}). At a given
  $\Sigmacl$, large, massive clouds have smaller $f_{\rm esc}^{\rm cum}(3\Myr)$
  than compact, less massive clouds owing to longer evolutionary timescales.
  Dense, cluster-forming clumps that are destroyed within $3 \Myr$ ({\tt
    M1E4R03}, {\tt M1E4R02}, and {\tt M1E5R05}) have relatively high values of
  $f_{\rm esc}^{\rm cum}(3\Myr)\sim 40\%$--$50\%$.

\item {\textit{Solid-Angle Weighted Optical Depth PDF}}

  For an isotropic point source of radiation, the escape fraction is determined
  by the solid-angle weighted PDF $P_{\Omega}$ of the optical depth as seen from
  the point source through Equation~\eqref{eq:solid_pdf}. Assuming that
  $P_{\Omega}$ is lognormal, we demonstrate that the point-source escape
  fraction $\langle e^{-\tauc} \rangle_{\Omega}$ is much higher than would be
  estimated based on the solid-angle averaged optical depth, ($e^{-\taucmean}$),
  if the PDF has a large dispersion $\sigmac$ (Figure~\ref{f:lognorm}). We
  calculate $P_{\Omega}$ in our simulations by assuming that all radiation is
  emitted from the luminosity center of source particles. The shape of
  $P_{\Omega}$ is in general not lognormal, with peaks and dips associated with
  dense, star-forming clumps and photoevaporated, outflowing gas (top row of
  Figure~\ref{f:pdf}). Nevertheless, the lognormal estimates based on the mean
  and standard deviation of $P_{\Omega}$ are quite close to the true escape
  fraction $\langle e^{-\tauc} \rangle_{\Omega}$ from the luminosity center
  (Figure~\ref{f:fesc-lognorm}). We define the reduction factor
  $\mathcal{F} = -\ln (\langle e^{-\tauc} \rangle_{\Omega})/\taucmean$ that
  measures the effective optical depth relative to the mean value for
  non-ionizing radiation. We show that $\mathcal{F}$ decreases as $\taucmean$
  and/or $\sigmac$ increases (Figure~\ref{f:lognorm}).

\item {\textit{Area Weighted Optical Depth PDF}}

  We calculate the area-weighted PDF $P_{A}$ of the optical depth within the
  half-mass radius of gas, as would be measured by an external observer, finding
  that the shape and temporal change of $P_{A}$ are similar to those of the
  solid-angle PDF $P_{\Omega}$ (Figure~\ref{f:pdf}). Consistent with results
  from the solid-angle PDF, a simple estimate of the escape fraction
  ($e^{-\langle \tauext_{\rm n}\rangle_{A}}$) based on the area-averaged optical
  depth $\langle\tauext\rangle_{A}$ significantly underestimates the real escape
  fraction (Figure~\ref{f:fescn-tauh}). This is because
  $\langle\tauext\rangle_{A}$ is higher than $\taucmean$ measured from the
  luminosity center (due to path length differences) and does not properly
  account for the variance in optical depth; the latter is more important at
  high optical depth. The reduction in the effective optical depth tends is
  quite dramatic for larger $\langle \tauext_{\rm n}\rangle_A$
  (Figure~\ref{f:Fred2}), such that the effective optical depth is $\sim 1$ when
  $\langle\tauext\rangle_{A} \sim 10$.

  We present two simple methods for estimating $\fescn$ for observed
  star-forming regions. In the first method, we assume a marginally resolved
  cloud for which only the area-averaged dust optical depth
  $\langle\tauext\rangle_{A}$ is observationally available. We show that our
  results agree with the estimate of the escape fraction
  $\fescn^{\rm est,1} = e^{-\eta_1\langle\tauext\rangle_{A}}$, where the
  correction factor $\eta_1 = 0.56/(1 + 1.25\langle\tauext\rangle_{A}^{0.48})$
  depends only on $\langle\tauext\rangle_{A}$. In the second method, we assume
  that the area distribution $P_A$ of the dust optical depth $\tauext_{\rm n}$
  surrounding sources is observationally available. We show that our results
  agree well with the estimate based on an area average,
  $\fescn^{\rm est,2} = \langle e^{-\eta_2 \tau^{\rm ext}_{\rm n}}\rangle_A$
  with a constant correction factor $\eta_2=0.3$. The two methods both yield
  estimates within $\sim 20\%$ of the actual luminosity-weighted escape fraction
  obtained from adaptive ray tracing in our simulations
  (Figure~\ref{f:fesc_pred}).

\end{enumerate}

\subsection{Discussion}\label{s:discuss}


\subsubsection{Comparison with Other Simulations}\label{sec:comparison}

In previous work, we compared radiation fields computed from a two-moment
radiation scheme with $M_1$-closure relation with those computed with ART for
identical distributions of sources and gas density, and found that the two
methods are in good agreement with each other in terms of large-scale radiation
field and escape fraction \citepalias{kim17}. One would therefore expect similar
results for $f_{\rm esc,n}$ to the findings reported in \citet{ras17}, which
used the $M_1$ scheme to study the interaction between non-ionizing radiation
and gas using the same basic cloud model as we adopt here \citep[see][but note
specific model parameters differ]{ras16}. In practice, however, it is not
meaningful to make a detailed comparison because the evolution of SFE with time
diverges between simulations that use ART and those that use M1.
\citetalias{kim18} compared our ART simulations with the results of
\citet{ras16} and showed that use of the $M_1$ method can overestimate the SFE,
since radiation forces are underestimated in the vicinity of star
particles.\footnote{Although this may be ameliorated by specialized local
  treatment \citep[][]{ros15} when there is a single point source, the accuracy
  of the $M_1$ solution is necessarily limited in regions with multiple
  radiation sources.} Thus, while trends of cumulative $f_{\rm esc, n}$ with
cloud properties are quite similar here to those reported in \citet{ras17},
specific models cannot be directly compared.

It is even more difficult to make comparisons of escape fractions with other
simulations in which not just the radiative transfer scheme but also cloud
parameters, treatment of sink/source particles, dust opacity, and feedback
mechanisms are quite different from those we have considered. Nevertheless, it
is noteworthy to observe that there is a consistent common trend among different
studies of decreasing cumulative (or instantaneous) LyC escape fraction at the
time of the first supernovae with increasing cloud mass. In our simulations,
low-mass clouds that evolve rapidly ($\tffcl < 1\Myr$) and are destroyed before
$t' = 3\Myr$ have $\fescicum(3\Myr) \gtrsim 0.4$, while massive
($\Mcl = 10^6\Msun$) clouds with $\tffcl \gtrsim 5\Myr$ have
$\fescicum(3\Myr) \sim 0.1$ (see Section~\ref{s:fesccum} and
Table~\ref{t:result}). Likewise, \citet{dal12} found that dense, compact clouds
(their Runs F, I, J) exhibit $\fesci \gtrsim 0.8$ at the time of the first
supernovae, whereas massive ($\Mcl = 10^6 \Msun$) clouds (their Runs A, B, X)
have $\fesci \lesssim 0.2$. \citet{kimm19} report that the luminosity-weighted,
time-averaged escape fraction is only $5.2\%$ for a solar-metallicity cloud with
$\Mcl = 10^6\Msun$, $\Sigmacl \sim 1.3 \times 10^2\Sunit$, and $M_* = 10^5\Msun$
over the cloud lifetime $20 \Myr$. In \citet{how18}, the cumulative escape
fraction of LyC radiation at $t'= 5\Myr$ is only $8\%$ for a cloud with
$10^6 \Msun$ and $\Sigmacl \sim 280\Sunit$, while less massive
($\Mcl = 5\times 10^4$, $10^5\Msun$) clouds with $\Sigmacl \sim 10^2\Sunit$ are
almost entirely destroyed before $t' = 5\Myr$ and have
$\fescicum(5\Myr) \sim 0.64$. Taken together, these results suggest that the
escape of radiation before the time of the first supernovae is intimately linked
to the timescale of cloud evolution.

As noted in Section~\ref{s:evol}, several other groups observed (as did we) an
overall monotonic increase of LyC escape fraction $\fesci$ with time in their
simulations, as an increasing fraction of photons escapes through low-density
channels created by feedback \citep[e.g.,][]{wal12,dal13,kimm19}. In contrast,
\citet{how18} found large fluctuations (up to a factor of $\sim 6$) in $\fesci$
over short ($\lesssim 1\Myr$) timescales as small-scale turbulent flows around
sources absorb photons and make \HII\ regions ``flicker''. While it is difficult
to fully ascertain the causes of the difference, it is likely to reflect
different subgrid models for star formation and/or radiation-gas interaction.
For example, \citet{how18} assume that only a fraction of gas mass accreted onto
a sink particle is converted into stars. The remaining gas in the ``reservoir''
would lower the light-to-mass ratio of the sink particle and make \HII\ regions
become more easily trapped by accretion flows.

In addition to affecting the short-term evolution of $\fesc$, ``subgrid''
treatment of radiation in the immediate vicinity of star particles can also
affect the local collapse and therefore the cumulative star formation efficiency
and escape fraction for different RHD methods or subgrid model treatments, as
recently emphasized by \citet[][]{kru18} and \citet{hop19}. We investigate some
aspects of this question in Appendix~\ref{s:appendixA} by exploring differing
subgrid models for local escape fractions. Our conclusion is that provided the
resolution is sufficiently high, effects on cloud evolution (and therefore
$\fesc$) are relatively modest.

\subsubsection{Implications for Diffuse Ionized Gas and Galaxy-Scale Escape
  Fraction}

Based on work summarized in Section~\ref{s:intro}, ionizing radiation from young
massive stars is the only known source that can explain the maintenance of
diffuse ionized gas in the Galaxy and in external galaxies \citep{haf09}. This
relies on a substantial fraction of ionizing photons escaping from natal clouds,
but direct evidence of this escape has been lacking. In our simulations, the
cumulative escape fraction of ionizing photons before the onset of supernova
feedback in GMCs with typical gas mass $\Mcl \sim 10^5\Msun$ and surface density
$\Sigmacl \sim 10^2 \Sunit$ is $30$--$40\%$ (Figure~\ref{f:fcum_Sigma}). This
suggests that a substantial fraction of UV photons produced by massive stars can
escape into the surrounding ISM through low-density holes induced by turbulence
and radiation feedback. Our work thus supports the claim that leakage of
ionizing photons from \HII\ regions is responsible for the photoionization of
the warm ionized medium in the diffuse ISM.

Understanding how stellar ionizing photons can leak out of host galaxy's ISM and
make it all the way to the intergalactic medium is still under active
investigation \citep[e.g.,][]{wis14,ma15,paa15,kimm17,kak19,rig19,mcc19}.
Observational studies that directly detect escaping Lyman continuum radiation
indicate that the LyC escape fraction is generally small with
$\fesci^{\rm gal} \sim 1$--$10\%$ or less
\citep[e.g.,][]{lei13,bor14,izo16,lei16}, with only a few exceptions
\citep[e.g.,][]{sha16,izo18,riv19}. Unless a galaxy is completely obscured by
dust, the difference between $\fesci^{\rm gal}$ and $\fescn^{\rm gal}$ is
expected to be large. This is in contrast to the similarity between cloud-scale
escape fraction $\fesci$ and $\fescn$ found in the present work, which we
interpret as being due to the high ionization parameter (or low hydrogen neutral
fraction) in classical \ion{H}{2} regions, which makes dust grains the primary
absorber of both ionizing and non-ionizing radiation
(Section~\ref{s:similarity}). Due primarily to the geometric dilution of
radiation, the diffuse ionized gas exhibits line ratios characteristic of gas in
a low stage of ionization (e.g., [\ion{S}{2}] and [\ion{N}{2}]) and low
ionization parameter \citep{dom94, mat00, sem00}. This suggests that neutral
hydrogen absorption is more important in diffuse ionized gas than in \HII\
regions and $\fesci^{\rm gal}$ would be reduced more relative to
$\fescn^{\rm gal}$. The preliminary results for the galaxy-scale escape
fractions obtained by post-processing galactic disk simulations with adaptive
ray tracing are indeed in agreement with this expectation (Kado Fong et al.,
2019 in preparation).

\subsubsection{Implications for Cloud-Scale Star Formation Indicators}

The escape of a substantial fraction of UV photons (both ionizing and
non-ionizing) from \HII\ regions also has important implications for
observational determinations of star formation rates and efficiencies on cloud
scales. Most star formation rate estimators are tied to the luminosity from
massive stars, with optical emission lines such as H$\alpha$ from photoionized
gas being the most traditional indicators. However, rate indicators based on
H$\alpha$ emission (or free-free emission, also produced by photoionized gas)
cannot fully recover the intrinsic ionizing luminosity of a cluster because of
dust absorption \citep[e.g.,][]{bin18}. For this reason, combinations of
H$\alpha$ (or UV) and IR measurements have been extensively explored to
calibrate the dust absorption (as well as dust attenuation of recombination
emission lines) and are widely adopted in Galactic and extragalactic studies
\citep[see][for review]{ken12}. Unfortunately, calibration to account for the
escape of radiation has been largely ignored. This is likely not a serious issue
for measuring large-scale star formation, assuming the ISM overall acts as a
bolometer \citep[but see][]{hec11}. However, for individual star-forming clouds
the use of star formation rate indicators correcting only for dust absorption
may systematically underestimate the true star formation rate, considering that
the escape fraction of radiation may be appreciable.

In this regard, our proposed methods for estimating $\fescn$ (see Section
\ref{s:PDF-pred} for $\fescn^{\rm est,1}$ and $\fescn^{\rm est,2}$) can be
useful for recovering the bolometric luminosity of star clusters. The column
density distribution of Galactic molecular clouds has been extensively studied
using CO line emission \citep[e.g.,][]{goo09}, near-IR dust extinction
\citep[e.g.,][]{kai09}, and far-IR thermal dust emission maps
\citep[e.g.,][]{lom14}. For star-forming clouds that are well resolved, the
distribution of observed optical depth with a correction factor can be used to
directly estimate
$\fescn \approx \langle \exp(-0.3\tau_{\rm n}^{\rm ext})\rangle_A$; this also
provides an upper bound on $\fesci$. In cases where the overall size of the
cloud can be measured but the column density distribution is unavailable due to
poor resolution (presumably for most massive star-forming clouds in external
galaxies), one may utilize the area-averaged dust optical depth, again applying
a correction factor, with
$\fescn \approx \exp\left(\tfrac{-0.56\langle\tau_{\rm n}^{\rm ext}\rangle_A}{1
    + 1.25\langle\tau_{\rm n}^{\rm ext}\rangle_A^{0.48}}\right)$. When applied
to our simulation data, these methods approximate the actual escape fraction to
within $\sim 20\%$ (Figure~\ref{f:fesc_pred}).

\subsubsection{Potential Effect of Dust Destruction}\label{s:ddest}

Our simulation results suggest that dust absorption plays an important role in
controlling the escape fraction of radiation. While we adopted constant dust
absorption cross sections for both ionizing and non-ionizing radiation, dust
grains (e.g., small carbon grains and PAHs) in \HII\ regions can be destroyed by
intense UV radiation field
\citep[e.g.,][]{voi92,tie08,deh10,lop14,sal16,bin18,cha19}. This can potentially
lead to an increase in the escape fraction. To study this question
quantitatively, we have run additional models for the fiducial cloud in which
dust grains absorbing ionizing radiation (and non-ionizing radiation) are
completely destroyed in fully ionized gas, full details of which can be found in
Appendix~\ref{s:appendixB}. Our results suggest that, although the overall cloud
evolution is quite similar to the standard model without dust destruction, the
boost in escape fraction can be significant. Under the assumption that ionizing
radiation is not absorbed by dust in ionized regions, we find cumulative
$\fescicum (t^{\prime} = 3 \Myr) = 0.5$, which is 0.2 higher than the standard
model and close to the value \citet{kimm19} found ($\sim 0.5$--$0.6$) for their
model {\tt M5\_SFE10}, which is fairly similar in cloud parameters and SFE to
our model. Since the complete destruction of dust in ionized gas is unlikely to
occur in reality, our results put an upper limit on the escape of radiation in
\HII\ regions with dust destruction. Ideally, future models should incorporate
the effects of varying grain properties that depend on the local radiative and
chemical environment \citep[e.g.,][]{gla19} to provide more realistic estimate
of dust absorption and escape fractions.

\subsubsection{Limitations of the Current Model}

Finally, we comment on the potentially important effects of physical processes
that are not modeled in our simulations. Our simulations neglect
radiation-matter interaction at subgrid scales, adopting $f_{\rm esc,*}=1$ from
sink particle regions. Since we neglect potential small-scale absorption
\citep[e.g.,][]{kru18}, the cloud-scale escape fraction that we calculate may be
an overestimate. However, our simulations also neglect other forms of
pre-supernovae feedback such as stellar winds and/or protostellar outflows,
which in principle could can further increase the porosity of the gas
surrounding sources and increase the escape of radiation from cloud scales. In a
low-metallicity environment, radiation pressure exerted by resonantly scattered
Lyman-$\alpha$ photons can play an important role in disrupting clouds and
raising the escape fraction \citep{kimm19}. After $t=t_{*,0} + 3 \Myr$,
supernovae explosions of most massive stars occurring inside molecular clouds
may effectively clear out the remaining gas and increase the escape fraction of
radiation \citep[e.g.,][]{rog13,gee15,iff15}. This is particularly the case for
massive clouds whose evolutionary timescale is expected to be longer than
$3\Myr$ \citepalias{kim18}. However, the cumulative escape fraction of ionizing
radiation may not increase significantly due to a sharp drop in the photon
production rate caused by the death of the massive stars
\citep[e.g.,][]{kimm14,kimm19}. Expansion of superbubbles driven by multiple
supernovae may further help UV photons propagate hundreds of parsecs from the
birth cloud \citep[e.g.,][]{dov00,kimcg17a,tre17}. Multi-scale simulations of
GMC evolution with comprehensive feedback mechanisms included are necessary to
fully understand the escape of UV radiation in realistic environments. As a
first step towards this goal, efforts to incorporate UV radiation feedback in
the TIGRESS numerical framework \citep{kimcg17b}, which models a local patch of
galactic disk with self-consistent star formation and supernovae feedback, are
currently underway.

\acknowledgements The authors thank the anonymous referees for their helpful
comments and suggestions. J.-G.K. thanks Bruce Draine for sharing his thoughts
on dust destruction in ionized gas. J.-G.K. was supported by Lyman Spitzer Jr.
Postdoctoral Fellowship (Princeton University) and the National Research
Foundation of Korea (NRF) through the grant NRF-2014-Fostering Core Leaders of
the Future Basic Science Program. The work of W.-T.K. was supported by the
National Research Foundation of Korea (NRF) grant funded by the Korea government
(MSIT) (2019R1A2C1004857). This work was also supported by the U.~S. National
Science Foundation under grant AST-1713949 to E.~C.~O. The computation of this
work was supported by the Supercomputing Center/Korea Institute of Science and
Technology Information with supercomputing resources including technical support
(KSC-2017-C3-0029), and the PICSciE TIGRESS High Performance Computing Center at
Princeton University.

\software{{\tt Athena} \citep{sto08}, {\tt yt} \citep{tur11}, {\tt numpy}
  \citep{van11}, {\tt matplotlib} \citep{hun07}, {\tt IPython} \citep{per07},
  {\tt pandas} \citep{mck10}.}

\appendix

\section{Effect on Star Formation Efficiency of Escape Fraction at Subgrid
  Scale}\label{s:appendixA}

\subsection{Background}

For simulations that include point sources of stellar radiation and gravity, it
is important to treat radiation pressure and gravitational forces at small
scales consistently. In our simulations, the gravitational force produced by
star particles is computed using a particle-mesh scheme \citep{gon13}. This
approach inevitably loses its accuracy in the vicinity of sink particles as each
point mass is smeared out over a few grid cells over which the
gas distribution is
unresolved. The radiation force is calculated using the volume-averaged
radiation flux returned from the ray tracing \citepalias{kim17}, but it also
suffers the momentum cancellation effect on the smallest resolved scales. It has
been proposed that momentum associated with radiation forces can be injected
under the assumption of an isotropic radial flux distribution at subgrid scales
\citep{hop19}. However, for consistency one would also have to include subgrid
gravity, and the proper subgrid treatment of extinction is unclear.

To avoid potentially inconsistent treatment of gravity and radiation for regions
surrounding a sink/source particle, we assume that all the gas accreted onto the
boundary faces of the $3^3$-cell control volume does not interact with
radiation, falls without obstruction, and is instantly converted into stars. We
also allow photons emitted by a source to interact with gas only after they
emerge from its control volume \citepalias{kim18}. The complete neglect of gas-radiation
interactions inside the unresolved control volume results in the maximal gas
accretion and radiation escape in our models: the corresponding escape fraction
is $f_{\rm esc,*}=1$ from unresolved scales. Physically, this situation
would hold in the limiting case
when accreting gas is extremely clumpy with a negligible covering fraction.

Recently, \citet{kru18} considered the opposite extreme in which accreting
material is smooth and spherically symmetric and interacts with radiation most
strongly. He showed that a steady-state spherical inflow solution does not exist
if the outward radiation force (by both UV and dust-reprocessed IR radiation)
exceeds the inflow momentum rate (by inward gravity), which occurs when the mass
inflow rate $\dot{M}_*$ is smaller than a critical rate
$\dot{M}_{\rm *,crit,sph}$. Conversely, in spherical symmetry radiation is
completely smothered by the accreting flows if
$\dot{M}_* > \dot{M}_{\rm *,crit,sph}$. In the absence of IR radiation (and
photoionization), the critical accretion rate is given by
$\dot{M}_{\rm *,crit,sph} = {L_*}/{[cv_{\rm in}(r_{\rm s})]}$, where $L_*$ is
the source luminosity and $v_{\rm in}(r_{\rm s}) = \sqrt{2G M_*/r_{\rm s}}$ is
the free-fall inflow speed at the dust sublimation radius
$r_{\rm s} \approx 3.4 \times 10^2\,{\rm AU} (L_*/10^6 \Lsun)^{1/2}$ for the
dust sublimation temperature $T_{\rm s} = 1500\Kel$. \citet{kru18} also showed
through 1D numerical experiments that the numerical resolution should be high
enough to resolve the dust sublimation radius to successfully reproduce the
behavior of the spherical steady-state solution. In unresolved runs, the radius
at which photon momentum is deposited moves outward so that radiation
incorrectly overcomes the inward momentum of the inflowing material and can
freely escape. As the dust sublimation radius is practically impossible to
resolve for simulations of star clusters or galaxies, he proposed a subgrid
model for radiation feedback in which the emergent luminosity on the resolved
scale is $L_*$ (i.e., $f_{\rm esc,*}=1$) if
$\dot{M}_* < \dot{M}_{*,{\rm crit,sph}}$, and zero ($f_{\rm esc,*}=0$)
otherwise.

Since accreting material is likely clumpy, but has a nonzero covering fraction,
the reality should be somewhere in between the two extremes discussed above.
However, the question of how strongly infalling material interacts with
radiation and modifies the emergent radiation field is an interesting and
complicated problem on its own; understanding these processes would require
numerical simulations resolving a wide range of spatial scales. In this
Appendix, we instead conduct a simple experiment to explore the impact of
varying the grid-scale escape fraction $f_{\rm esc,*}$ on the cloud-scale SFE.

\subsection{Subgrid Model}

We make following assumptions about accreting flows and radiation-matter
interactions at subgrid scales.

\begin{enumerate}
\item A gaseous parcel accreting onto the control volume of a sink particle
  continues to move inward without being turned around by the radiation pressure
  force. Accretion of a parcel into the sink region would occur if its surface
  density exceeds the Eddington surface density
  $\Sigma_{\rm Edd} = \Psi/(4\pi c G) = 370\Sunit (\Psi/10^3\Lsun\Msun^{-1})$
  \citep[e.g.,][]{ras17}. Because surface densities naturally increase in the
  converging inflow, parcels that exceed $\Sigma_{\rm Edd}$ at resolved scales
  would be expected to grow in surface density at smaller scales. With $\Sigma$
  ever larger than $\Sigma_{\rm Edd}$, these parcels would therefore continue to
  accrete inward to unresolved scales. We note that because material arrives at
  the sink region in an inhomogeneous state, we do not consider it necessary for
  the accretion rate to exceed $\dot{M}_{\rm *,crit,sph}$ in order to
  successfully overcome the effects of radiation; this differs from
  \citet{kru18}, which assumed spherical accretion at all distances.

\item For a steady-state free-falling inflow, the angle-averaged density profile
  is $\langle\rho(r)\rangle_{\Omega} = \dot{M}_*/[4\pi r^2 v_{\rm in}(r)]$ for
  $v_{\rm in}(r)= (2G M_*/r)^{1/2}$. The solid angle-averaged dust optical depth
  from $r_{\rm s}$ to the edge of the control volume
  $\Delta x \;(\gg r_{\rm s})$ is then given by
\begin{align}
  \langle \tau_{\rm s}\rangle_{\Omega}
  & = \int_{r_{\rm s}}^{\Delta x} \kappad
    \langle \rho (r)\rangle_{\Omega} dr \nonumber \\
  & \equiv \beta_{\rm p}\kappad\Sigma_{\rm Edd} =
    76 \beta_{\rm p} \left(\tfrac{\kappad}{500
    \cm^2\gram^{-1}}\right) \left(\tfrac{\Psi}{10^3
    \Lsun\Msun^{-1}}\right)\,,\label{e:taus}
\end{align}
where
\begin{align}\label{e:betap}
  \beta_{\rm p} & = \dfrac{\dot{M}_*}{\dot{M}_{*,{\rm crit,sph}}} \\
                & = 3.5 \left(\tfrac{\dot{M}_*}{10^{-3}\Msun\yr^{-1}}\right)
                  \left(\tfrac{\Psi}{10^3 \Lsun\Msun^{-1}}\right)^{5/4}
                  \left(\tfrac{M_*}{10^3\Msun}\right)^{3/4}
\end{align}
is a dimensionless parameter characterizing the inflow rate of momentum due to
the free-falling gas normalized to the rate of momentum injection by stellar
radiation.

\item The instantaneous escape fraction for a sink particle is set to
\begin{equation}\label{e:subgrid}
  f_{\rm esc,*} = \exp(-\mathcal{F}_* \langle\tau_{\rm s}\rangle_{\Omega})\,,
\end{equation}
where the reduction factor $0 \le \mathcal{F}_* \le 1$ is a free parameter
characterizing the clumpiness of the inflowing gas at subgrid scales. When
$\mathcal{F}_* = 0$, gas-radiation interactions are negligible inside the
control volume, while $\mathcal{F}_* = 1$ corresponds to a spherically
symmetric inflow.
\end{enumerate}

\begin{figure}[htb!]
  \epsscale{1.15}\plotone{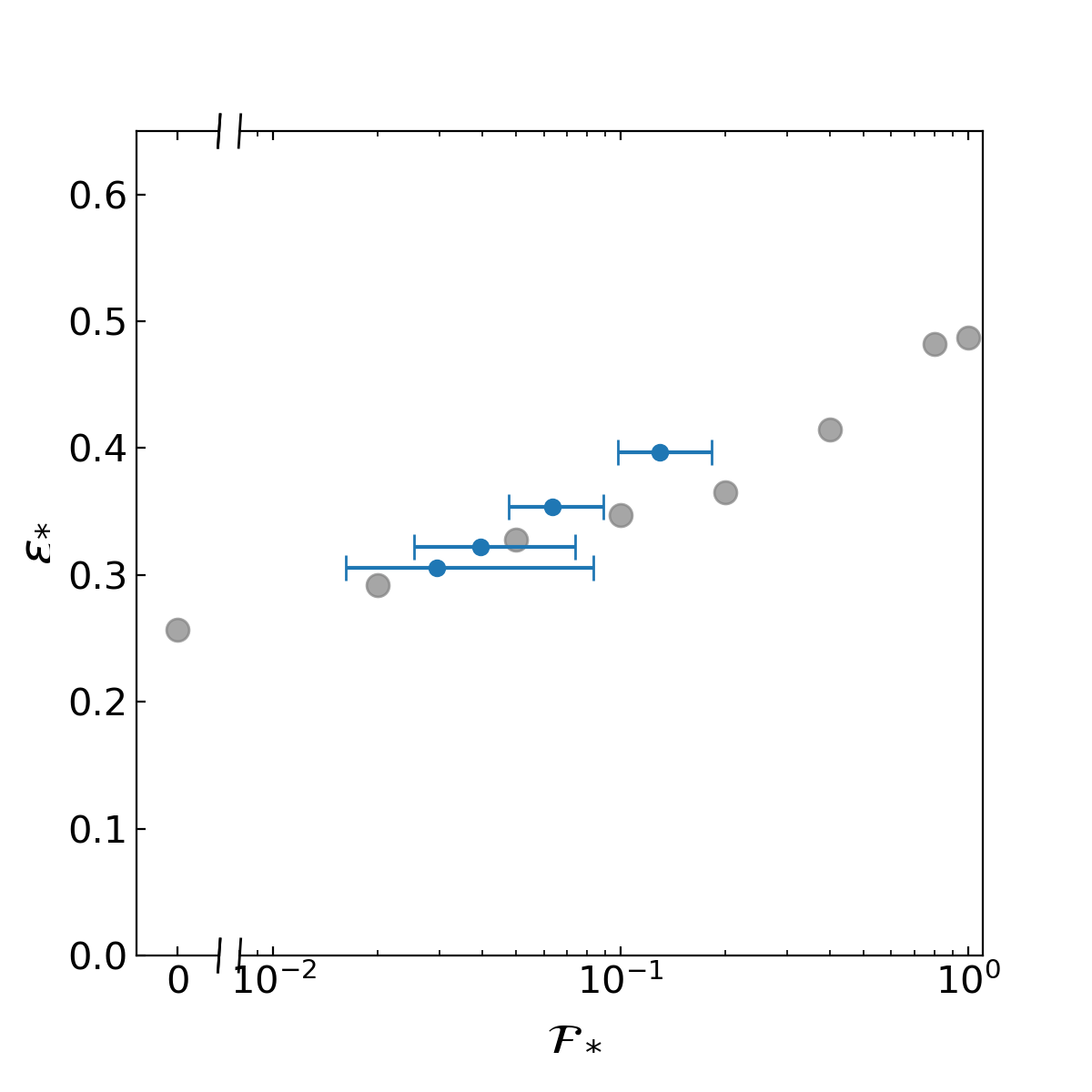}
  \caption{Net SFE from the radiation pressure-only simulations of the fiducial
    cloud with $\Mcl = 10^5 \Msun$ and $\Rcl = 20\pc$ as a function of the
    subgrid-scale reduction factor $\mathcal{F}_*$. The subgrid-scale escape
    fraction of non-ionizing radiation is set to
    $f_{\rm esc,*} = \exp(-\mathcal{F}_*\langle \tau_{\rm s}\rangle_\Omega)$,
    where $\langle \tau_{\rm s}\rangle_\Omega \propto \dot{M}_*$ is the solid
    angle-averaged dust optical depth from the dust sublimation radius to the
    resolved scale $\Delta x$. The gray circles are from the models with fixed
    $\mathcal{F}_*$. The blue circles with horizontal bars are from the runs in
    which $\mathcal{F}_*$ is set to vary according to the mass inflow rate 
    under the assumption that the optical-depth PDFs are lognormal with standard
    deviation $\sigmac = 1.0, 1.5, 2.0, 2.5$ from right to left. The horizontal
    position of the bars and blue circles mark 25th, 50th, 75th percentiles of the
    $\mathcal{F}_*$ distribution when
    $10^{-6}\Msun\yr^{-1} < \dot{M}_* < 10^{-2}\Msun\yr^{-1}$.}\label{f:SFE}
\end{figure}

\subsection{Effect on Star Formation Efficiency}

We perform a set of numerical simulations by adopting Equation \eqref{e:subgrid}
with fixed $\mathcal{F}_*$ for the grid-scale escape fraction of individual sink
particles. We consider the fiducial cloud ($\Mcl = 10^5 \Msun$ and
$\Rcl = 20\pc$) with radiation pressure feedback only (no photoionization).
Figure~\ref{f:SFE} plots as gray circles the net SFE from the models with
various $\mathcal{F}_*$. The net SFE increases mildly from 0.26 to 0.49 as
$\mathcal{F}_*$ varies from 0 to 1. This is because as the feedback efficiency
decreases (with increasing $\mathcal{F}_*$), more stellar mass is required to
drive outflows to disrupt the cloud. We find that the timescales for star
formation and cloud destruction do not vary by more than 10\% in all runs with
differing $\mathcal{F}_*$.

\begin{figure}[!t]
  \epsscale{1.2}\plotone{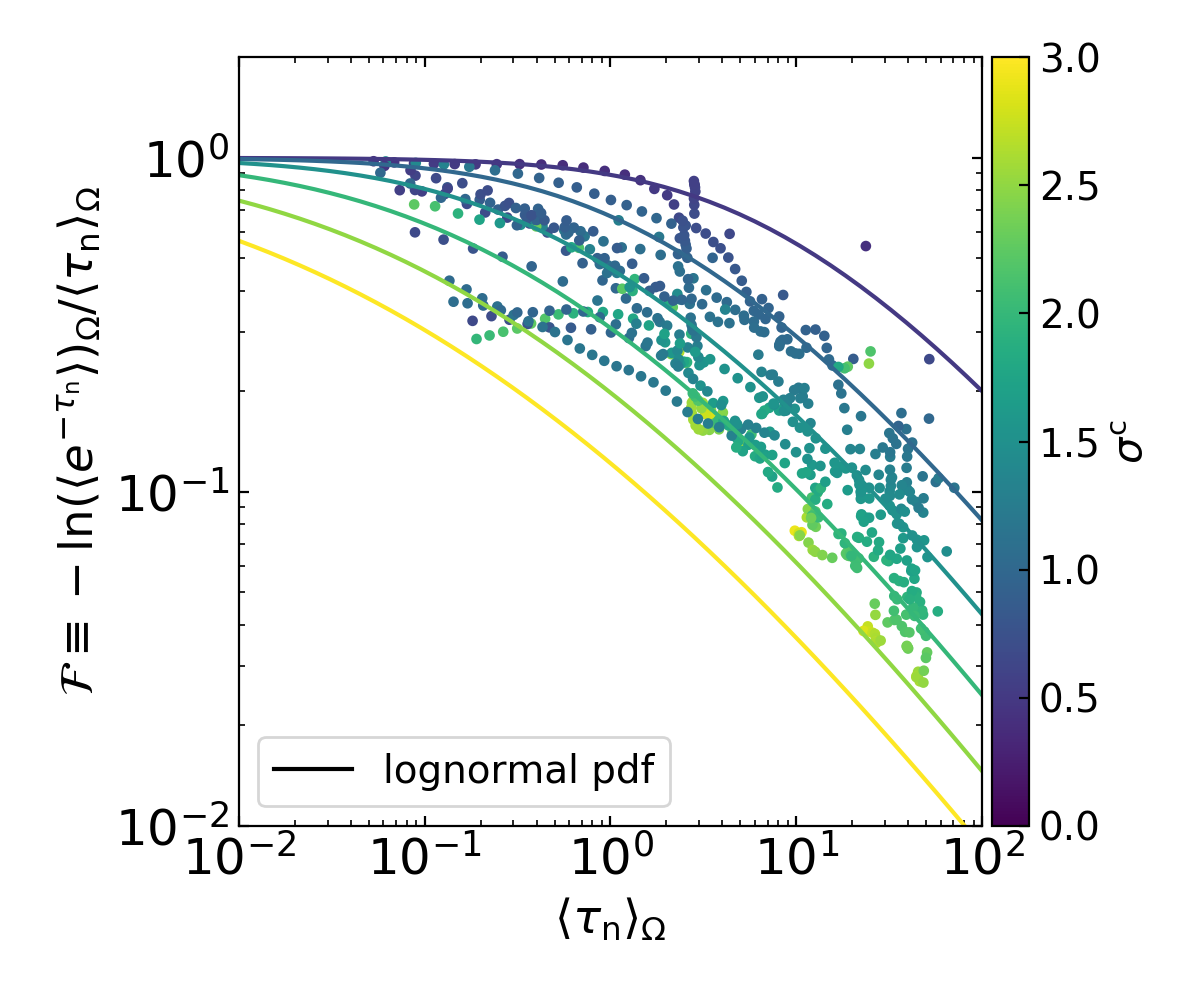}
  \caption{Cloud-scale reduction factor $\mathcal{F}$ for non-ionizing radiation
    as measured from the stellar center of luminosity as a function of
    $\langle \tau_{\rm n}^{\rm c} \rangle_{\Omega}$. The circles are from our
    simulations, with colors corresponding to $\sigmac$. The lines draw the
    reduction factor expected for lognormal PDFs with
    $\sigmac = 0.5, 1, \cdots, 3.0$ from top to bottom.}\label{f:fred-alt}
\end{figure}

We also consider models in which $\mathcal{F}_*$ varies over time consistently
with the instantaneous sink luminosity and accretion rate. While we cannot
directly determine $\mathcal{F}_*$ without running simulations that resolve
structures close to $r_{\rm s}$, the cloud-scale reduction factor $\mathcal{F}$
based on the {\it resolved} radiation-gas interaction
(Section~\ref{s:PDF-Omega}) can shed some light on the possible behavior of
$\mathcal{F}_*$ at subgrid scales. Figure~\ref{f:lognorm}(b) showed that
$\mathcal{F}$ tends to decrease with the width $\sigma_c$ of the optical-depth
PDF and the solid angle-averaged optical depth $\taucmean$.
Figure~\ref{f:fred-alt} plots $\mathcal{F}$ again, this time as a function of
$\taucmean$. Most of the simulation results (filled circles) are consistent with
the predictions of lognormal PDFs (lines) with
$1.0\lesssim \sigmac \lesssim 2.5$. In the absence of photoionization and other
feedback processes, one may assume that the accreting flows remain clumpy at
subgrid scales and the reduction factor $\mathcal{F}_*$ is correlated with the
angle-averaged optical depth of the accreting flow
$\langle\tau_{\rm s}\rangle_{\Omega}$ in the same way that $\mathcal{F}$ is
correlated with $\taucmean$ as displayed in Figure~\ref{f:fred-alt}.

\begin{figure*}[!t]
  \epsscale{0.95}\plotone{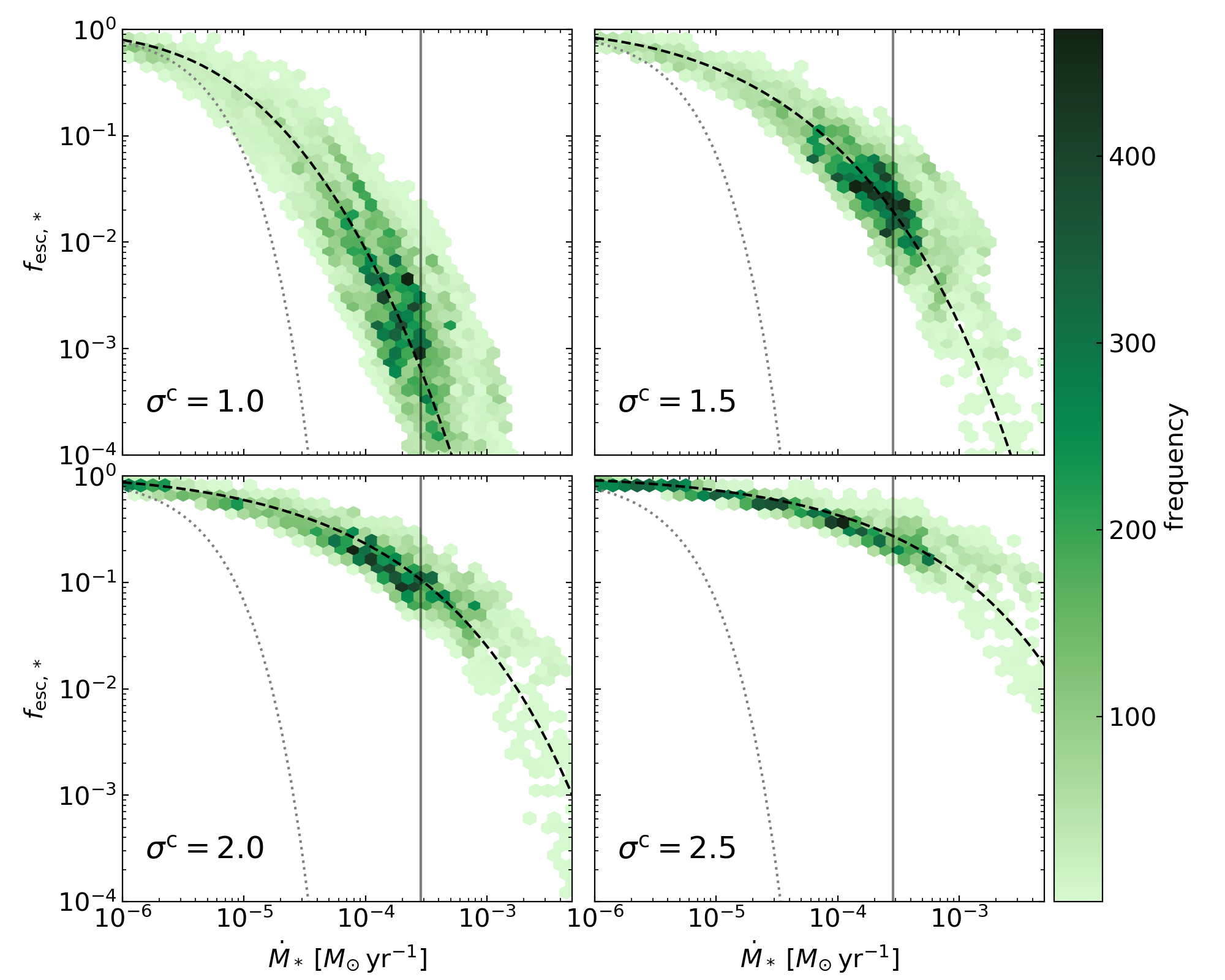}
  \caption{2D histograms of the grid-scale escape fraction $f_{\rm esc,*}$ and
    the mass accretion rate $\dot{M}_*$ onto sink particles for the fiducial
    cloud model, under the assumption that the optical-depth PDF at subgrid
    scales follows a lognormal distribution with the mean
    $\langle \ln \tau_{\rm s} \rangle_{\Omega} = \ln (\langle\tau_{\rm
      s}\rangle_{\Omega}) + (\sigmac)^2/2$ and the standard deviation
    $\sigmac=1.0, 2.0, 2.0, 2.5$. The vertical solid lines mark the critical
    mass accretion rate $\dot{M}_{\rm *,crit,sph}$ for the existence of steady
    spherical inflow solutions \citep{kru18}. The thick dashed lines draw the
    relationship between $\dot{M}_*$ and $f_{\rm esc,*}$ expected for a cluster
    particle with mass $M_* = 10^3\Msun$ and a constant light-to-mass ratio
    $\Psi = 10^3 \Lsun\Msun^{-1}$, while the thin dotted lines correspond to the
    case of the spherical accretion with
    $\mathcal{F}_*=1$.}\label{f:fesc-subgrid}
\end{figure*}

We run four simulations assuming that the optical-depth PDF on subgrid scales
follows a lognormal distribution with $\sigmac = 1.0$, $1.5$, $2.0$, and $2.5$.
In each run with given $\sigmac$, we calculate
$\langle\tau_{\rm s}\rangle_{\Omega}$ for individual sink particles using
Equation \eqref{e:taus}, obtain $\mathcal{F}_*$ from
$\langle\tau_{\rm s}\rangle_{\Omega}$ using the corresponding
$\mathcal{F}$--$\taucmean$ relation shown as solid lines in
Figure~\ref{f:fred-alt}, and then apply the instantaneous escape fraction
according to Equation \eqref{e:subgrid}.

Figure~\ref{f:fesc-subgrid} plots the 2D histograms on the
$f_{\rm esc,*}$--$\dot{M}_*$ plane of sink particles in all runs. For reference,
the thick dashed curves plot the relationships between $\dot{M}_*$ and
$f_{\rm esc,*}$ expected for a cluster particle with mass $M_* = 10^3\Msun$ and
light-to-mass ratio $\Psi=10^3 \Lsun\Msun^{-1}$ from
Equations~\eqref{e:taus}--\eqref{e:subgrid}. Even when $\mathcal{F}_*$ is
allowed to vary with $\langle\tau_{\rm s}\rangle_{\Omega}$, the escape fraction
still decreases with increasing $\dot{M}_*$, but much more mildly compared to
the case of spherical accretion with $\mathcal{F}_*=1$, plotted as thin dotted
lines. When $\sigmac = 2.0$, for example, the grid scale escape fraction
$f_{\rm esc,*}$ drops below $\sim 10\%$ if
$\dot{M}_* \gtrsim \dot{M}_{\rm *,crit,sph} = 2.8 \times 10^{-4}\Msun\yr^{-1}
(M_*/10^3\Msun)^{3/4}$. However, the mass accretion phase with
$\dot{M}_* > \dot{M}_{\rm *,crit,sph}$ and $f_{\rm esc,*} \ll 1$ lasts only for
a brief period of time ($\lesssim 1\Myr$), so that the evolution of cloud-scale
escape fraction is not significantly affected by the choice of $\sigmac$. The
median values of $\mathcal{F}_*$ for
$10^{-6}\Msun\yr^{-1} < \dot{M}_* < 10^{-2}\Msun\yr^{-1}$ in the four runs with
$\sigmac = (1.0, 1.5, 2.0, 2.5)$ are (0.13, 0.063, 0.039, 0.029), respectively.
The resulting relationship between the net SFE and the subgrid-scale escape
fraction from all four runs is plotted in Figure \ref{f:SFE} as blue circles
with horizontal bars, which is overall similar to the results with fixed
$\mathcal{F}_*$.

To conclude, our numerical experiments demonstrate that the grid-scale escape
fraction has a modest impact on the effectiveness of radiation feedback in
halting accretion and controlling the SFE. To better understand the detailed
process of gas-radiation interactions and provide constraints on $\mathcal{F}_*$
on smaller scales, it is desirable to perform numerical simulations of accretion
flows onto massive stars resolving a wide range of spatial scales in the
presence of various feedback mechanisms.

\begin{figure}[!t]
  \epsscale{1.0}\plotone{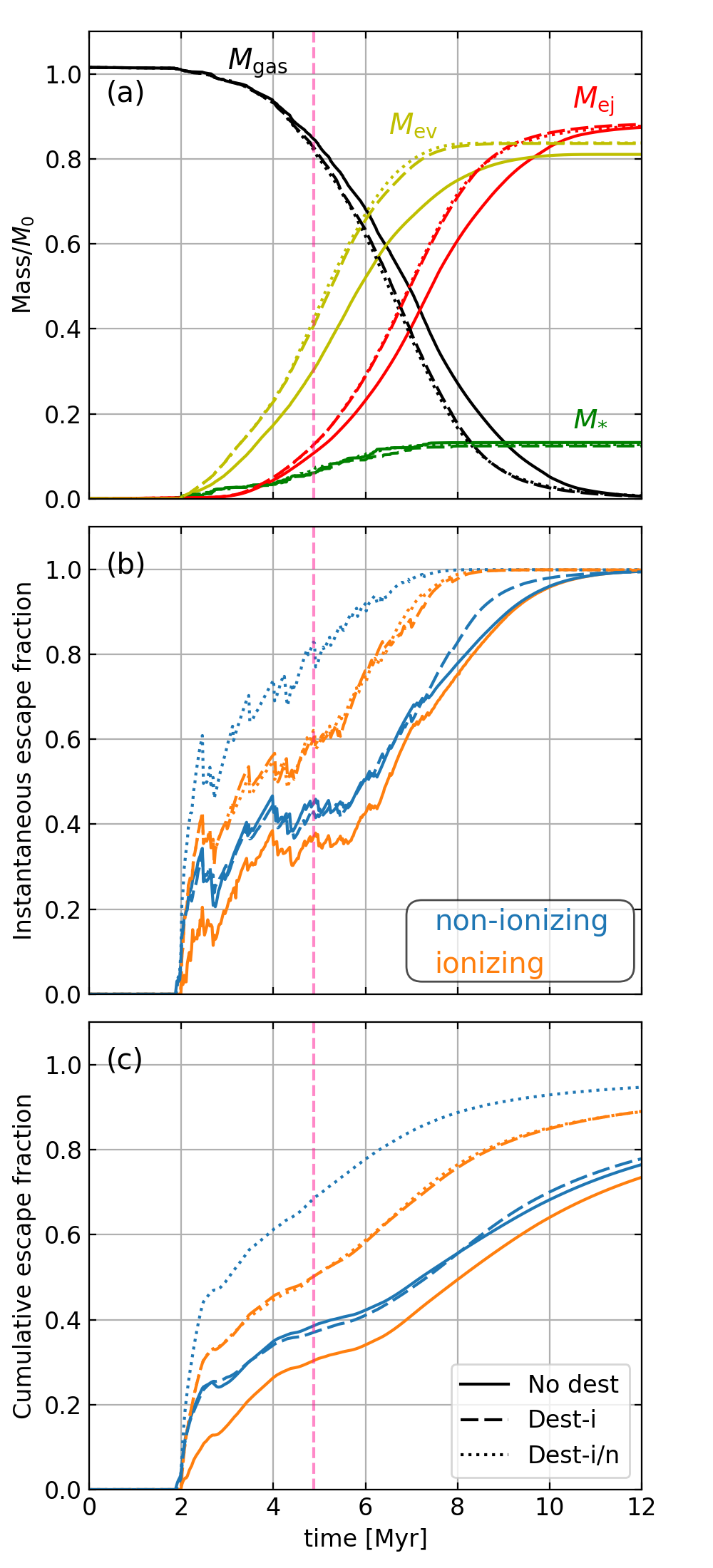}
  \caption{Effect of dust destruction in ionized gas for the fiducial cloud with
    $\Mcl = 10^5\Msun$ and $\Rcl = 20\pc$. Time evolution of (a) total gas mass
    in the simulation domain (black), stellar mass (green), ejected gas mass
    (red), and photoevaporated gas mass (yellow); (b) instantaneous and (c)
    cumulative escape fractions for ionizing (orange) and non-ionizing (blue)
    radiation. The solid lines show the standard model (``No dest'') in which
    dust absorption cross section $\sigma_{\rm d}$ is constant everywhere
    ($1.17\times 10^{-21}{\rm cm}^2\,{\rm H}^{-1}$). The dashed lines show the
    comparison run in which destruction of dust grains in ionized gas is assumed
    to prevent absorption of ionizing radiation ({\tt Dest-i}), while the dotted
    lines show the comparison run in which destruction of dust grains in ionized
    gas is assumed to prevent absorption of both non-ionizing and ionizing
    radiation ({\tt Dest-i/n}). The vertical dashed line incidates the epoch of
    the first supernova ($t^{\prime} = 3\Myr$).}\label{f:dust_dest}
\end{figure}

\section{Effects of Dust Destruction for the Fiducial Model}\label{s:appendixB}

Our simulations adopted a constant grain absorption cross section
$\sigma_{\rm d,i} = \sigma_{\rm d,n} = 1.17\times 10^{-21}{\rm cm}^2\,{\rm
  H}^{-1}$ for both ionizing and non-ionizing radiation. For dust models
characteristic of the diffuse ISM \citep[e.g.,][]{wei01}, the UV extinction
steeply rises toward shorter wavelengths and peaks at $h\nu \sim 17\eV$ due
mainly to small carbon grains and PAHs \citep{dra03,gla19}; the
frequency-averaged dust absorption cross section for ionizing radiation is
$\sim (1$--$1.5) \times 10^{-21} \cm^2\,{\rm H}^{-1}$ for ionizing stars with
blackbody temperature $2.5 \times 10^4 \Kel < T_* < 5.0 \times 10^4 \Kel$ and
has a
slightly lower ($\sim 20$--$30\%$) value at FUV wavelengths \citep{dra11b}.
However, both theory and observations suggest that dust grains are destroyed by
intense UV radiation in star-forming clouds
\citep[e.g.,][]{voi92,tie08,deh10,lop14,sal16,bin18,cha19}, which may lessen the
UV extinction and boost escape of radiation in ionized gas. For example,
\citet{gla19} adopted \citet{wei01}'s case A size distributions
and found that $\sigma_{\rm d,i}$ can be reduced by a factor $\sim2$--3 if the two
log-normal components representing PAHs and very small carbon grains
($b_{\rm C}=0.0$) are completely absent.

To examine the potential impact of dust destruction on the escape of radiation,
we have run additional simulations for the fiducial model ($\Mcl = 10^5 \Msun$,
$\Rcl = 20\pc$, $N=256$) assuming that dust grains are destroyed in ionized gas.
Since dust properties in \HII\ regions are quite uncertain, we consider two
extreme situations to bracket the range of possible outcomes:
\begin{itemize}
\item model {\tt Dest-i}: the dust absorption cross section for ionizing radiation
  scales with the neutral fraction
  ($\sigma_{\rm d,i} = \xn \times 1.17\times 10^{-21}{\rm cm}^2\,{\rm H}^{-1}$),
  but the cross section for non-ionizing radiation remains unchanged.
\item model {\tt Dest-i/n}: the dust absorption cross section for both ionizing and
  non-ionizing radiation scales with the neutral fraction
  ($\sigma_{\rm d,i} = \sigma_{\rm d,n} = \xn \times 1.17\times 10^{-21}{\rm
    cm}^2\,{\rm H}^{-1}$).
\end{itemize}
The first of these corresponds to the preferential destruction of
grains that absorb ionizing radiation in ionized gas, and the second
corresponds to destruction of all grains in ionized gas.  A similar
approach has been adopted by \citet{lau09}, \citet{how17}, and \citet{kimm19}. Since         
the complete destruction of dust grains in \HII\ regions is unlikely
to occur in reality,
these models put an upper limit on the escape of radiation in \HII\ regions.

Figure~\ref{f:dust_dest}(a) plots the temporal evolution of gas mass
($M_{\rm gas}$, black), stellar mass ($M_*$, green), ejected gas mass
($M_{\rm ej}$, red), and photoevaporated gas mass ($M_{\rm ev}$, yellow) for
different models. In the absence of dust grains to absorb ionizing photons in
\HII\ regions ($\fdusti \approx 0$), we expect photoionization feedback to
be more efficient and the outflow driving by radiation pressure to be less
efficient or absent (e.g., radiation pressure exerts no force on ionized gas in
model {\tt Dest-i/n}). As photoionization is the dominant feedback mechanism in low
surface-density clouds \citepalias{kim18}, it has a greater impact on
the simulation outcome than radiation pressure. 
However, the overall cloud evolution does not change much.
Compared to the standard model, the photoevaporation efficiency
($M_{\rm ev,final}/\Mcl$) is only 0.03 higher, while the net SFE is lower by
less than 0.01 in both models {\tt Dest-i} and {\tt Dest-i/n}.

In contrast, the evolution of the escape of radiation is noticeably different
from the standard model. Figure~\ref{f:dust_dest}(b) and (c) show the evolution
of instantaneous and cumulative escape fractions for ionizing (orange) and
non-ionizing (blue) radiation. At $t^{\prime} = 3\Myr$, the instantaneous escape
fractions are $(\fesci, \fescn) = (0.60, 0.42)$ in model {\tt Dest-i} and
$(\fesci, \fescn) = (0.62, 0.83)$ in model {\tt Dest-i/n}, while the cumulative
escape fractions are $(\fescicum, \fescncum) = (0.50, 0.37)$ in model {\tt Dest-i}
and $(\fescicum, \fescncum) = (0.50, 0.69)$ in model {\tt Dest-i/n}.
In model {\tt Dest-i},
$\fescn$ and $\fescncum$ evolve quite similarly to those in the standard model,
but are smaller than $\fesci$ and $\fescicum$. In both models {\tt Dest-i} and {\tt
  Dest-i/n}, $\fescicum$ at $t^{\prime} = 3\Myr$ is 0.20 greater than that in
the standard run ($\fescicum = 0.30$). In the standard model,
ionizing photons escape through
low-density density-bounded sightlines along which absorption by neutral
hydrogen is unimportant and the escape fraction is mainly determined by dust
absorption (Section~\ref{s:similarity}). Therefore, the escape fraction $\fesci$
in the absence of dust absorption simply reflects (the luminosity-weighted
average of) the fraction of the sightlines that are density-bounded.


\bibliographystyle{apj_hyperref}


\end{document}